\documentstyle[12pt]{article}
\begin{document}
\setlength{\oddsidemargin}{-1cm}
\setlength{\evensidemargin}{-1cm}
\newcommand{\nc}{\newcommand}
\nc{\beq}{\begin{equation}}
\nc{\eeq}{\end{equation}}
\nc{\bea}{\begin{eqnarray}}
\nc{\eea}{\end{eqnarray}}
\nc{\ba}{\begin{array}}
\nc{\ea}{\end{array}}
\nc{\nn}{\nonumber}
\nc{\bpi}{\begin{picture}}
\nc{\epi}{\end{picture}}
\nc{\scs}{\scriptstyle}
\nc{\sss}{\scriptscriptstyle}
\nc{\sst}{\scriptstyle}
\nc{\ts}{\textstyle}
\nc{\ds}{\displaystyle}
\nc{\sctn}[1]{\section{\hspace{-18pt}.\ #1}}
\nc{\subsctn}[1]{\subsection{\hspace{-16pt}.\ #1}}
\nc{\subsubsctn}[1]{\subsubsection{\hspace{-14pt}.\ #1}}

\nc{\al}{\alpha}
\nc{\be}{\beta}
\nc{\ga}{\gamma}
\nc{\Ga}{\Gamma}
\nc{\de}{\delta}
\nc{\De}{\Delta}
\nc{\ep}{\epsilon}
\nc{\ve}{\varepsilon}
\nc{\eb}{\bar{\eta}}
\nc{\et}{\eta}
\nc{\ka}{\kappa}
\nc{\la}{\lambda}
\nc{\La}{\Lambda}
\nc{\th}{\theta}
\nc{\Th}{\Theta}
\nc{\ze}{\zeta}
\nc{\p}{\partial}

\nc{\cw}{c_\th}
\nc{\oocw}{\cw^{-1}}
\nc{\cwcw}{\cw^2}
\nc{\cwcwcw}{\cw^3}
\nc{\cwcwcwcw}{\cw^4}
\nc{\oocwcw}{\cw^{-2}}
\nc{\sw}{s_\th}
\nc{\oosw}{\sw^{-1}}
\nc{\swsw}{\sw^2}
\nc{\swswsw}{\sw^3}
\nc{\swswswsw}{\sw^4}
\nc{\tw}{t_\th}
\nc{\twtw}{\tw^2}

\nc{\mb}[1]{\mbox{#1}}
\nc{\unit}{\mb{\bf\large 1}}
\nc{\Tr}{\mb{Tr}}
\nc{\Bb}{\mb{\boldmath $\ds B$}}
\nc{\Wb}{\mb{\boldmath $\ds W$}}
\nc{\Xb}{\mb{\boldmath $\ds X$}}
\nc{\Fb}{\mb{\boldmath $\ds F$}}
\nc{\Tb}{\mb{\boldmath $\ds T$}}
\nc{\Vb}{\mb{\boldmath $\ds V$}}
\nc{\mw}{m_{\sss W0}}
\nc{\mz}{m_{\sss Z0}}
\nc{\half}{{\ts\frac{1}{2}}}
\nc{\ihalf}{{\ts\frac{i}{2}}}
\nc{\dg}{\dagger}
\nc{\Lag}{{\cal L}}
\nc{\Lgf}{\Lag_{\mbox{\scriptsize\em gf}}}
\nc{\Lgh}{\Lag_{\mbox{\scriptsize\em gh}}}
\nc{\Lgfgh}{\Lag_{\mbox{\scriptsize\em gf,gh}}}
\nc{\Lwhc}{\Lag_{\mbox{\scriptsize\em Whc}}}
\nc{\Lbhc}{\Lag_{\mbox{\scriptsize\em Bhc}}}
\nc{\Lvhc}{\Lag_{\mbox{\scriptsize\em Vhc}}}
\nc{\Luhc}{\Lag_{\mbox{\scriptsize\em Uhc}}}
\nc{\od}{{\cal O}}
\nc{\mubar}{\bar{\mu}}
\nc{\Laeff}{\La_{\rm eff}}

\nc{\ltap}{\;\raisebox{-.4ex}{\rlap{$\sim$}}\raisebox{.4ex}{$<$}\;}
\nc{\gtap}{\;\raisebox{-.4ex}{\rlap{$\sim$}}\raisebox{.4ex}{$>$}\;}

\begin{flushright}
\normalsize
Freiburg-THEP 97/17\\
August 1997\vspace{0.8cm}\\
\end{flushright}

\begin{center}
{\Large\bf Corrections to oblique parameters induced by\vspace{0.2cm}\\
anomalous vector boson couplings}\vspace{1.2cm}\\
J.J.\ van der Bij and Boris Kastening\vspace{0.4cm}\\
\it Albert-Ludwigs-Universit\"at Freiburg\\
\it Fakult\"at f\"ur Physik\\
\it Hermann-Herder-Stra\ss e 3\\
\it D-79104 Freiburg\\
\it Germany\vspace{1.2cm}\\
\end{center}

\begin{center}
{\bf Abstract}\vspace{0.4cm}\\
\begin{minipage}{16cm}
We study quadratically divergent radiative corrections to the oblique
parameters at CERN LEP1 induced by non-standard vector boson self-couplings.
We work in the St\"uckelberg formalism and regulate the divergences
through a gauge-invariant higher derivative scheme.
Using consistency arguments together with the data we find a limit
on the anomalous magnetic moment $\De \ka$ of the W-boson,
$|\De\ka|\ltap 0.26$.
\end{minipage}
\vspace{0.8cm}\\
\end{center}

\sctn{Introduction}
With the running of the CERN $e^+e^-$ collider LEP-200 and with results from
the Fermilab Tevatron the self-interactions of the vector bosons are nowadays
being measured directly.
Within the standard model the vector boson self-interactions are fully
determined by the gauge structure of the theory.
Deviations from the standard model can be parametrized by a set of operators
describing so-called anomalous couplings and experiment can put a limit on the
coefficient of these operators.
However the presence of anomalous gauge boson self-couplings will violate the
renormalizability of the theory.
As a consequence one can generate divergent contributions to quantities at
lower energies than the two vector boson threshold.
When one uses a cut-off procedure one can estimate the induced effects and use
low-energy data to put limits on the assumed anomalous couplings.
As the data at low energy have become very precise since LEP-100,
strong limits can be found.
Indeed in a recent discussion it was argued that the LEP-100 data can
obviate the LEP-200 data, with the exception of so-called blind directions
in coupling constant space.
These blind directions correspond to operators that do not have direct
effects in propagators and can therefore only be seen after insertion
inside a loop, indirectly generating propagator effects.
In the original articles \cite{orig,Bi1} on these induced effects
quadratically and quartically divergent contributions were found, leading
to relatively severe restrictions.
These results were criticized in \cite{BuLo,EiWu}, where it was
argued that the quadratic divergences would be gauge-dependent and
non-physical, so one should use dimensional regularization as a cut-off,
which gives logarithmic divergence and weak constraints.
In a more recent calculation \cite{HeVe},€dimensional regularization
in $d$ dimensions was used to determine the quadratic divergences as
poles in $(d-2)$.
In \cite{HaIs..} the divergences were regularized by using the Higgs field
as a regulator.
An analysis based on the philosophy of \cite{BuLo,EiWu} is presented
in \cite{DaVa}.
Both calculations \cite{HeVe,HaIs..} confirm the original calculations as
having quadratic divergent contributions, as is consistent with power counting
in chiral perturbation theory.
However also here the situation is not fully satisfactory, as only one
cut-off scale is assumed to be present.
In reality however, there are different cut-off scales present.
This is most easily seen from the vector boson propagators, which consist of
longitudinal and transversal parts, which could have different form factors. 
Indeed one would expect the longitudinal part to have structure at a
relatively low scale, as this part describes effects coming from the
Goldstone boson part of the theory, dependent on the mechanism of
spontaneous symmetry breaking, where strong interactions might be present.
In order to clarify the situation we therefore perform in this paper a
calculation of induced low energy effects from anomalous effects using a
higher derivative regulator.
More precisely, we describe vector boson physics without a Higgs boson as a
gauged non-linear sigma-model.
The anomalous couplings are then given by higher dimensional operators.
This is the St\"uckelberg formalism and is closely related to chiral
perturbation theory.
This has the advantage that the whole calculation can be performed in a
gauge-invariant way.
The quadratic and higher divergences are regulated via covariant higher
derivative terms; the remaining logarithmic ones via dimensional
regularization.

We limit the anomalous couplings to terms that have no CP-violation,
as we know CP-violation to be very small.
Furthermore we limit the discussion to terms that correspond to dimension
four operators in the unitary gauge. 
Within the standard model there is an extra custodial $SU(2)_R$
symmetry in the limit of vanishing hypercharge, which has as a consequence
that the $\rho$-parameter deviates from unity only through hypercharge
couplings.
This symmetry has to be protected at least to some level also in the
anomalous couplings and we will focus mostly on the operators where the
custodial symmetry is only violated through a minimal coupling to
hypercharge.

We have assumed, that the only relevant gauge bosons are those of the
$SU(2)_L\times U(1)_Y$ gauge group and that new physics does not couple
directly to light fermions.
Therefore any contribution of new physics below the vector-boson pair
threshold can only come from vacuum polarization corrections to gauge
boson propagators \cite{PeTa,MaBuLo}.
For most of the available low-energy, $Z$ and $W$ observables it is
possible to parametrize these corrections by the six so-called oblique
parameters $S$, $T$, $U$, $V$, $W$, $X$ \cite{PeTa,MaBuLo}.
These parameters are therefore well suited to compare experiment with our
calculation and we will use a recent analysis in this terminology.
 
In section \ref{model}, we will outline the model we use to describe
the electroweak sector of the standard model, give the various anomalous
couplings and describe our regularization procedure.
In section \ref{oblpar}, we give our results for the oblique parameters.
In section \ref{validity} we investigate the contribution of our
regularization procedure to the oblique parameters and study the consistency
of the method.
In section \ref{exp}, we analyze our results with respect to experimental
data.

\sctn{The Model}
\label{model}
Since the origin of electroweak symmetry breaking is unknown, we
do not assume the existence of a Higgs field, but describe the
breaking using the St\"uckelberg formalism \cite{St,BaDaVa}.
That is we write the spontaneously broken $SU(2)_L\times U(1)_Y$ theory
as a gauged non-linear sigma model.

We need the following definitions. 
Let
\beq
\Wb_{\mu\nu}=\half\tau_aW_{\mu\nu}^a
=\p_\mu\Wb_\nu-\p_\nu\Wb_\mu+ig[\Wb_\mu,\Wb_\nu]
\eeq
and
\beq
\Bb_{\mu\nu}=\half\tau_3B_{\mu\nu}
=\p_\mu\Bb_\nu-\p_\nu\Bb_\mu
\eeq
be the $SU(2)_L$ and $U(1)_Y$ field strengths.
Let $U$ be an $SU(2)$ valued field that describes the longitudinal
degrees of freedom of the vector fields and let $U$ transform as
\beq
U\rightarrow U_LUU_Y
\eeq
under $SU(2)_L\times U(1)_Y$ gauge transformations with
$U_L=\exp(-\ihalf g\vec{\Th}_L\cdot\vec{\tau})$
and
$U_Y=\exp(-\ihalf g'\Th_Y\tau_3)$, where $g'$ is the hypercharge coupling.
Define auxiliary quantities
\beq
\ts\Tb=U\tau_3U^\dg
\eeq
and 
\beq
\ts\Vb_\mu=-\frac{i}{g}(D_\mu U)U^\dg
\eeq
with
\beq
D_\mu U=\p_\mu U+ig\Wb_\mu U+ig'U\Bb_\mu\,.
\eeq
Under $SU(2)_L\times U(1)_Y$ gauge transformations, they transform as
$\Tb\rightarrow U_L\Tb U_L^\dg$ and
$\Vb_\mu\rightarrow U_L\Vb_\mu U_L^\dg$.

Electroweak theory without fermions and without the Higgs scalar is
then described by the Lagrange density
\beq
\label{lagew}
\ts\Lag_{EW}=
-\half\Tr(\Wb_{\mu\nu}\Wb^{\mu\nu})
-\half\Tr(\Bb_{\mu\nu}\Bb^{\mu\nu})
+\frac{g^2v^2}{4}\Tr(\Vb_\mu \Vb^\mu)\,,
\eeq
where $v$ replaces the vacuum expectation value of the Higgs field.

In this formalism, the $CP$ conserving anomalous three and four vector
boson couplings that are of dimension four in unitary gauge ($U=\unit$)
are described by the following set of gauge-invariant operators
\bea
\label{ano1}
\Lag_1&=&-i\Tr(\Wb^{\mu\nu}[\Vb_\mu,\Vb_\nu])\,,\\
\Lag_2&=&-\ihalf B^{\mu\nu}\Tr(\Tb[\Vb_\mu,\Vb_\nu])\,,\\
\Lag_3&=&-\ihalf\Tr(\Tb\Wb^{\mu\nu})\Tr(\Tb[\Vb_\mu,\Vb_\nu])\,,\\
\Lag_4&=&(\Tr[\Vb_\mu\Vb_\nu])^2\,,\\
\Lag_5&=&(\Tr[\Vb^\mu\Vb_\mu])^2\,,\\
\Lag_6&=&\Tr(\Vb^\mu\Vb^\nu)\Tr(\Tb\Vb_\mu)\Tr(\Tb\Vb_\nu)\,,\\
\Lag_7&=&\Tr(\Vb^\mu\Vb_\mu)(\Tr[\Tb\Vb_\nu])^2\,,\\
\Lag_8&=&(\Tr[\Tb\Vb_\mu])^2(\Tr[\Tb\Vb_\nu])^2\,,
\label{ano8}
\eea
which we introduce by adding
\beq
\Lag_{ano}=\sum_{i=1}^8 g_i\Lag_i
\eeq
to $\Lag_{EW}$.
In our treatment, the aforementioned approximate custodial $SU(2)_R$
symmetry is realized by $U\rightarrow UU_R$ with $U_R\in SU(2)$.
Among the operators (\ref{ano1})-(\ref{ano8}), only $\Lag_1$, $\Lag_4$
and $\Lag_5$ conserve this custodial symmetry in the limit of vanishing
hypercharge coupling.
At the same time, the absence of the other operators leads to a
cancellation of quartic divergences in oblique electroweak parameters
\cite{Bi1}, as will be seen below.
In other words, $\Lag_1$, $\Lag_4$ and $\Lag_5$ correspond to the
so-called blind directions in coupling constant space which do not
receive the severe constraints that the presence of quartic divergences
would impose \cite{RuGa..}.
We will therefore assume that the custodial symmetry is respected by
the anomalous couplings and thus restrict our analysis with respect to
experimental results to these three operators.

Since higher than logarithmic divergences are set to zero by dimensional
regularization, we have to parametrize them using a different method.
We will apply the method of higher covariant derivatives \cite{FaSl}.
In the version used here, it leaves only logarithmic divergences
in the anomalous contribution to the oblique parameters in Landau gauge.
These remaining divergences are then regulated dimensionally.
Specifically, we add to the theory
\beq
\label{laghctr}
\ts\Lag_{hc,tr}
=\frac{1}{2\La_W^2}\Tr[(D_\al\Wb_{\mu\nu})(D^\al\Wb^{\mu\nu})]
+\frac{1}{2\La_B^2}\Tr[(\p_\al\Bb_{\mu\nu})(\p^\al\Bb^{\mu\nu})]
\eeq
for the transverse degrees of freedom of the gauge fields and
\beq
\label{laghclg}
\ts\Lag_{hc,lg}
=-\frac{g^2v^2}{4\La_V^2}\Tr[(D^\al\Vb^\mu)(D_\al\Vb_\mu)]
\eeq
for the longitudinal ones, where the $\La_X$ parametrize
the quadratic divergences and are expected to represent the
scales where new physics comes in.
The covariant derivatives in (\ref{laghctr}) and (\ref{laghclg}) are
defined by
\bea
D_\al\Wb_{\mu\nu}&=&\p_\al\Wb_{\mu\nu}+ig[\Wb_\al,\Wb_{\mu\nu}]\,,
\\
D_\al\Vb_\mu&=&\p_\al\Vb_\mu+ig[\Wb_\al,\Vb_\mu]\,.
\eea
As a variant of $\Lag_{hc,lg}$, one can use e.g.\
\bea
\Lag_{hc,lg}'
&=&\ts
-\frac{v^2}{4\La_V^2}\Tr[(D^\al D^\be U)(D_\al D_\be U)^\dg]
\nn\\
&=&\ts
-\frac{g^2v^2}{4\La_V^2}\left\{\Tr[(D^\al\Vb^\mu)(D_\al\Vb_\mu)]
+\frac{g}{2}\Lag_1+\frac{g'}{2}\Lag_2+g^2\Lag_4-\frac{g^2}{2}\Lag_5\right\}
\eea
instead, which is closer to a natural regularization in the linear model.
The quartic divergences are invariant under this change due to the
additional suppression factor $g^2v^2/\La_V^2$.
Once we impose absence of quartic divergences by setting
$g_2=g_3=g_6=g_7=g_8=0$, it is easily seen that now the quadratic
divergences are invariant under the change in regularization.

We remark here that a reasonable assumption of the dynamics would 
make $\La_V$ the smallest, being related to the Goldstone sector
of the theory.
Also, one would expect $\La_B$ to be very large, as it is hard
to imagine a fundamental dynamics, where strong interactions would start
in the Abelian sector of the theory.
The presence of the approximate custodial symmetry tells us that terms
with explicit $T$ or $B_{\mu \nu}$ should be heavily suppressed.
We finally note that the signs in front of $\La_V^2,\La_W^2,\La_B^2$
are not determined a priori.
The method of gauge fixing we use is outlined in appendix \ref{gf}.

Finally, our conventions lead to the following definitions of the usual
gauge fields:
\beq
\label{zafields}
\left(\ba{c}Z_\mu\\A_\mu\ea\right)
=\left(\ba{cc}c&s\\-s&c\ea\right)
\left(\ba{c}W_\mu^3\\B_\mu\ea\right)\,,
\eeq
\beq
\label{wfields}
W_\mu^\pm=\ts\frac{1}{\sqrt{2}}(W_\mu^1\mp iW_\mu^2)\,,
\eeq
where we have used the abbreviations $c=\cos\Th_W$, $s=\sin\Th_W$ and
where the weak mixing angle is defined by $\tan\Th_W=g'/g$.

\sctn{Oblique parameters}
\label{oblpar}
In models where new physics comes in at scales much larger than
the electroweak scale, it is usually assumed that an expansion
of the vacuum polarizations linear in $k^2$ is sufficiently
accurate to parametrize the new physics effects at the electroweak
scale.
Accordingly, a description of new physics effects in terms of three
parameters $S$, $T$, $U$ is appropriate \cite{PeTa}.
In our description this assumption is explicitly violated
as can be seen from the structure of the $k^2$ and $k^4$ terms
in the vacuum polarizations given in appendix \ref{g1g4g5} by
(\ref{piaa})-(\ref{piww}).
We therefore need all six parameters $S$, $T$, $U$, $V$, $W$, $X$
used when observables at the scales $0$, $m_Z^2$, $m_W^2$ are taken
into account and the above assumption is not valid \cite{MaBuLo}.

The six oblique parameters are computed from the $\Pi_{XY}^g(k^2)$
part of the non-Standard Model contribution to the vacuum polarizations,
\beq
\ts\Pi_{XY}^{\mu\nu}(k^2)
=\Pi_{XY}^g(k^2)g^{\mu\nu}+\Pi_{XY}^k(k^2)\frac{k^\mu k^\nu}{k^2}\,,
\eeq
with $XY=WW,ZZ,ZA,AA$.
Their definitions are, according to \cite{MaBuLo} (except that our
conventions lead to a different sign of $s$),
\bea
\label{Sdef}
\alpha S
&=&
4s^2 c^2\left[\frac{\Pi_{ZZ}^g(m_z^2)-\Pi_{ZZ}^g(0)}{m_z^2}
+\frac{c^2-s^2}{sc}\Pi_{ZA}^{g\prime}(0)
-\Pi_{AA}^{g\prime}(0)\right]\,,
\\
\label{Tdef}
\alpha T
&=&
\frac{\Pi_{WW}^g(0)}{m_w^2}-\frac{\Pi_{ZZ}^g(0)}{m_z^2}\,,
\\
\alpha U
&=&
4s^2\left[\frac{\Pi_{WW}^g(m_w^2)-\Pi_{WW}^g(0)}{m_w^2}
-c^2\frac{\Pi_{ZZ}^g(m_z^2)-\Pi_{ZZ}^g(0)}{m_z^2}
+2sc\Pi_{ZA}^{g\prime}(0)-s^2\Pi_{AA}^{g\prime}(0)\right]\,,
\\
\alpha V
&=&
\Pi_{ZZ}^{g\prime}(m_z^2)-\frac{\Pi_{ZZ}^g(m_z^2)-\Pi_{ZZ}^g(0)}{m_z^2}\,,
\\
\alpha W
&=&
\Pi_{WW}^{g\prime}(m_w^2)-\frac{\Pi_{WW}^g(m_w^2)-\Pi_{WW}^g(0)}{m_w^2}\,,
\\
\alpha X
\label{Xdef}
&=&
sc\left[\frac{\Pi_{ZA}^g(m_z^2)}{m_z^2}-\Pi_{ZA}^{g\prime}(0)\right]\,.
\eea
These combinations are well-suited for comparison with experimental
data.
In particular, the $W$ parameter only appears in the rather poorly
measured $W$ width and can therefore be dropped from the analysis.

To present the results of our calculation, a modification of the
$S$ and $U$ parameters is useful.
Let us define
\bea
\hat{S}&=&S-4s^2c^2V\,,\\
\hat{U}&=&U+4s^2c^2V-4s^2W\,.
\eea
In this way, $T$ is getting contributions only from $k$-independent terms,
$\hat{S}$ and $\hat{U}$ only from $k^2$ terms and $V$, $W$, $X$ only from
$k^4$ terms (higher powers of $k$ are absent in our treatment of the
quadratically divergent terms).
The relevance of this is that the $k^4$ terms of the various vacuum
polarizations are essentially identical, while our predictive power for the
$k^2$ terms and the $k$-independent terms hinges on additional assumptions,
as will be seen below.

Including all anomalous couplings (\ref{ano1})-(\ref{ano8}), we have
computed the quartically divergent contributions to the
$\Pi_{XY}^g(k^2)$.
The results can be found in appendix \ref{quartic}.
These contributions are $k$-independent.
A look at our definitions of the oblique parameters
(\ref{Sdef})-(\ref{Xdef}) shows that only $T$, representing
the correction to the $\rho$ parameter, depends on $k$-independent
parts of vacuum polarizations and therefore only it can be quartically
divergent.
We get
\bea
\label{Tquart}
\alpha T
&=&\ts
g_2^2\frac{\La_V^2\La_B^2}{\mw^4}\left(
-\frac{3}{4\ep}-\frac{5}{8}+\frac{3}{4}
\frac{\La_V^2\ln\frac{\La_V^2}{\mubar^2}
-\La_B^2\ln\frac{\La_B^2}{\mubar^2}}
{\La_V^2-\La_B^2}\right)
{+}(2g_1g_3{+}g_3^2)\frac{\La_V^2\La_W^2}{\mw^4}\left(
-\frac{3}{4\ep}-\frac{5}{8}+\frac{3}{4}
\frac{\La_V^2\ln\frac{\La_V^2}{\mubar^2}
-\La_W^2\ln\frac{\La_W^2}{\mubar^2}}
{\La_V^2-\La_W^2}\right)
\nn\\
&&\ts
+g_6\frac{\La_V^4}{\mw^4}
\left(-\frac{13}{4\ep}-\frac{31}{8}+\frac{13}{4}
\ln\frac{\La_V^2}{\mubar^2}\right)
+g_7\frac{\La_V^4}{\mw^4}
\left(-\frac{4}{\ep}-\frac{9}{2}+4\ln\frac{\La_V^2}{\mubar^2}\right)
+g_8\frac{\La_V^4}{\mw^4}
\left(-\frac{3}{\ep}-\frac{7}{2}
+3\ln\frac{\La_V^2}{\mubar^2}\right)
\nn\\
&&\ts
+\od(\La^2)\,.
\eea
Here, $\ep$ is defined by $d=4-2\ep$, where $d$ is the dimension
of spacetime.
From the presence of these quartic divergences we have therefore severe
constraints on the quartic vector boson couplings.
This is in agreement with \cite{Bi2}, but in contrast to \cite{BrEbGo},
who however use dimensional regularization and therefore find only a
logarithmic divergence. 
Evidently, absence of the custodial symmetry breaking couplings
$g_2$, $g_3$, $g_6$, $g_7$, $g_8$ leads to a cancellation of the
quartic divergencies in $T$.
In the further analysis we will therefore only keep the anomalous
couplings $g_1$, $g_4$, $g_5$ non-zero.
This is consistent with the dynamical principle from \cite{Bi1}, that
the breaking of the custodial symmetry should be only through the
minimal coupling to  hypercharge.

Our results for the $\Pi_{XY}^g(k^2)$ are given in appendix \ref{g1g4g5}.
From them we get
\bea
\label{Sres}
\alpha\hat{S}
&=&
-s^2\frac{g_1^2}{(4\pi)^2}
\left(\frac{20\La_V^2\La_W^4}{3\mw^2(\La_V^2-\La_W^2)^2}
+\frac{2\La_V^2\La_W^2(14\La_V^4-33\La_V^2\La_W^2+9\La_W^4)}
{3\mw^2(\La_V^2-\La_W^2)^3}
\ln\frac{\La_V^2}{\La_W^2}
\right)
\nn\\
&&
-2s^2\frac{\La_V^2}{\mw^2}\frac{g_1g}{(4\pi)^2}\left(\frac{1}{\ep}+1
-\ln\frac{\La_V^2}{\mubar^2}\right)\,,
\\
\label{Tres}
\alpha T
&=&
-\frac{3s^2}{4c^2}\frac{g_1^2}{(4\pi)^2}
\left(\frac{\La_B^2\La_W^2}{\mw^2(\La_B^2-\La_W^2)}
\ln\frac{\La_B^2}{\La_W^2}\right)
\nn\\
&&
+\frac{s^2\La_B^2}{c^2\mw^2}
\left[\frac{g_4}{(4\pi)^2}\left(
\frac{15}{4\ep}+\frac{13}{8}-\frac{15}{4}\ln\frac{\La_B^2}{\mubar^2}
\right)
+\frac{g_5}{(4\pi)^2}\left(
\frac{3}{2\ep}+\frac{5}{4}-\frac{3}{2}\ln\frac{\La_B^2}{\mubar^2}
\right)\right]\,,
\\
\label{Ures}
\alpha\hat{U}
&=&
\frac{2s^4}{c^2}\frac{g_1^2}{(4\pi)^2}
\frac{\La_V^2\La_B^2}{\mw^2(\La_V^2-\La_B^2)}
\ln\frac{\La_V^2}{\La_B^2}\,,
\\
\label{Vres}
\alpha V
&=&
-\frac{\La_V^2}{4\mw^2}\frac{g_1^2}{(4\pi)^2}\,,
\\
\label{Wres}
\alpha W
&=&
-\frac{\La_V^2}{4\mw^2}\frac{g_1^2}{(4\pi)^2}\,,
\\
\label{Xres}
\alpha X
&=&
\frac{s^2\La_V^2}{4\mw^2}\frac{g_1^2}{(4\pi)^2}
\eea 
for the quadratically divergent contributions to the oblique parameters.
The $1/\ep$ terms represent logarithmic divergences that are left
even after the quadratic divergences are parametrized by the scales
$\La_X$ and that do not cancel between the vacuum polarizations
in the oblique parameters.
Our interpretation is that the $1/\ep$ terms are replacing numerical
coefficients whose values depend on the details of what happens at the
scale where new physics comes in.

\sctn{Consistency of the Method}
\label{validity}
The results that we derived above cannot be compared directly with
experiment without some further considerations.
The reason for this is that the oblique corrections receive also
contributions from the regulator terms themselves and these contributions
should be consistent with the terms calculated from the radiative
corrections.

The tree-level contribution to the $\Pi_{XY}^g(k^2)$ can be read off
the quadratic part of the Lagrange density (\ref{lquad})
and is
\bea
\Pi_{AA}^g(k^2)
&=&
\left(\frac{s^2}{\La_W^2}+\frac{c^2}{\La_B^2}\right)k^4\,,
\\
\Pi_{ZA}^g(k^2)
&=&
sc\left(\frac{1}{\La_B^2}-\frac{1}{\La_W^2}\right)k^4\,,
\\
\Pi_{ZZ}^g(k^2)
&=&
\left(\frac{c^2}{\La_W^2}+\frac{s^2}{\La_B^2}\right)k^4
-\frac{\mz^2}{\La_V^2}k^2\,,
\\
\Pi_{WW}^g(k^2)
&=&
\frac{1}{\La_W^2}k^4-\frac{\mw^2}{\La_V^2}k^2\,.
\eea
The corresponding contributions to the oblique parameters are
\bea
\label{Stree}
\al S
&=&
4s^2\left(\frac{c^2}{\La_W^2}+\frac{s^2}{\La_B^2}
-\frac{1}{\La_V^2}\right)\mw^2\,,
\\
\label{Ttree}
\al T
&=&
0\,,
\\
\label{Utree}
\al U
&=&
4s^4\left(\frac{1}{\La_W^2}-\frac{1}{\La_B^2}\right)\mw^2\,,
\\
\label{Vtree}
\al V
&=&
\left(\frac{c^2}{\La_W^2}+\frac{s^2}{\La_B^2}\right)\frac{\mw^2}{c^2}\,,
\\
\label{Wtree}
\al W
&=&
\frac{\mw^2}{\La_W^2}\,,
\\
\label{Xtree}
\al X
&=&
s^2\left(\frac{1}{\La_B^2}-\frac{1}{\La_W^2}\right)\mw^2\,.
\eea
We observe that $\La_V$ enters only the $S$ parameter.

These tree-level contributions should be compared with the loop corrections
to check whether no inconsistency arises.
The philosophy we adopt here is the following.
The structure for the vector boson propagators, parametrized by
$\La_B$, $\La_W$, $\La_V$ is generated by the self-interactions
among the vector bosons, as parametrized by $g_1$.
Therefore the tree-level and the loop-corrections should be of similar size.
Whereas $S$, $T$, $U$ depend on the details of the interactions,
$V$, $W$, $X$ are given by a universal contribution.
We therefore impose the conditions $V_{tree} = V_{loop}$,
$W_{tree} = W_{loop}$, $X_{tree} = X_{loop}$.
This leads to the following result:
\bea
\label{labtree}
1/\La_B^2
&=&
0\,,
\\
\label{lawlavtree}
\frac{\mw^2}{\La_W^2}
&=&
- \frac{1}{4} \frac{g_1^2}{(4 \pi)^2}\frac{\La_V^2}{\mw^2}\,.
\eea
After imposing these conditions, consistency further demands that the radiative
corrections (\ref{Sres})-(\ref{Ures}) should be of the same order of magnitude
as the tree level relations (\ref{Stree})-(\ref{Utree}).
We see that this is indeed the case.
The relations (\ref{labtree}), (\ref{lawlavtree}) have an interesting physical
interpretation.
The fact that $\La_B\gg\La_W^2,\La_V^2$ means that the hypercharge field,
being a simple Abelian field, contains no structure.
Furthermore it is seen that the cut-off $\La_W$ is only an indirect
effect being generated by $g_1$, connected with the interactions in the
Goldstone boson sector.
Note the opposite signs for $\La_W^2$ and $\La_V^2$.
These relations were already qualitatively expected in section \ref{model}.
Given these relations, one can now make a comparison with experiment.

\sctn{Experimental Bounds}
\label{exp}
We use the following experimental constraints for oblique parameters,
which were provided to us by T.~Takeuchi.
They describe the deviation from standard model expectations for
$m_t=175GeV$, $m_H=300GeV$, $m_Z=91.18630GeV$,
$\al^{-1}=128.9$, $\al_S(m_Z)=0.123$:
\beq
\begin{tabular}{ll}
$S$&$= -1.0 \pm 1.5 $\,,\\
$T$&$= -0.57\pm 0.80$\,,\\
$U$&$=  0.07\pm 0.82$\,,\\
$V$&$=  0.49\pm 0.82$\,,\\
$X$&$=  0.22\pm 0.51$\,,
\end{tabular}
\eeq
with the correlation matrix
\beq
\label{corr}
\begin{tabular}{c|ccccc}
   &  $S$  &  $T$  &  $U$  &  $V$  &  $X$  \\ \hline
$S$&   1   &  0.79 &  0.54 & -0.77 & -0.95 \\
$T$&  0.79 &  1    & -0.05 & -0.98 & -0.56 \\
$U$&  0.54 & -0.05 &  1    &  0.05 & -0.76 \\
$V$& -0.77 & -0.98 &  0.05 &  1    &  0.55 \\
$X$& -0.95 & -0.56 & -0.76 &  0.55 &  1
\end{tabular}
\eeq
Although there is no Higgs particle in our model, the dependence
of the oblique parameters on the Higgs mass is very weak and we
can utilize the data above.
We will now use these data to put bounds on $\La_V$ and $\La_W$.
We will have to consider two cases, depending on the sign of $\La_V^2$.

\subsctn{The Case $\La_V^2 > 0$, $\La_W^2 < 0$}
In the comparison with experiment, we will now use the relations
(\ref{labtree}), (\ref{lawlavtree}) and give limits on $\La_W$ and
$\La_V$ from the formulae (\ref{Stree})-(\ref{Xtree}).
One might wonder whether it would not be more appropriate
to use formulae (\ref{Sres})-(\ref{Xres}), but here the comparison is
complicated due to the arbitrariness involved by the undetermined
coefficients.
The procedure we take gives the most conservative, i.e.\ the least restrictive
limits.
In order to facilitate the discussion, we change in this subsection the notation
$\La_W^2 \rightarrow - \La_W^2$.
We also define an auxiliary $\Laeff^2 = g_1^2 \La_V^2$.
We will use the data on $U$, $V$, $X$ to put a limit on $\La_W$.
Subsequently, we use the information on $S$ to put a limit on $\La_V$.

Using $U$, $V$ and $X$, we get from (\ref{corr}) the statistically
independent combinations
\bea
\label{dekalk2}
U-0.74V+2.0X &=& -0.14\pm 0.28\,,
\\
\label{dekalk3}
U-0.59V-0.72X &=& -0.4\pm 1.3\,,
\\
\label{dekalk4}
U+1.6V+0.087X &=& 0.9\pm 1.6\,,
\eea
which, using (\ref{Utree}), (\ref{Vtree}), (\ref{Xtree}),
(\ref{labtree}), (\ref{lawlavtree}), translate into
\beq
\Laeff^2=(0.4\pm 1.3)\mw^2\,,
\eeq
giving at 95\% confidence level
\beq
\Laeff^2<2.5\mw^2\,.
\eeq

Using (\ref{lawlavtree})and $\mw=80.26GeV$, this can be written as
\beq
\label{lawlimit}
\La_W > 1.3 TeV\,.
\eeq
Subsequently, using (\ref{Stree}), (\ref{labtree}) and the data on $S$,
we get at 95\% confidence level
\beq
\left(\frac {c^2}{\La_W^2(TeV)} + \frac{1}{\La_V^2(TeV)}\right) < 4.2\,.
\eeq
This can be written as 
\beq
\La_V > 0.49 TeV\,.
\eeq
When we express the results in terms of the anomalous magnetic moment
of the vector boson $\De \ka = g_1/g$ we get the following equation
\beq
\label{bound1}
|\De\ka| = \frac {0.25}{\La_W(TeV) \La_V(TeV)} \ltap 0.26\,.
\eeq
To arrive at the numerical bound, we took the linear combination
(\ref{dekalk3}), together with the bound on $S$ and their statistical
correlation, made a confidence level contour plot and determined the
value of $\De\ka$, where its line in the plot is tangential to the
ellipse bounded by $1.64\sigma$ lines.
Since we assume $\La_V^2 > 0$, $\La_W^2 < 0$, this gives an at least
$95\%$ confidence level bound on $\De\ka$ for this case.
This is a conservative procedure since it ignores some region in the plot
out side the $1.64\sigma$ ellipse that would also give smaller $|\De\ka|$.
Although (\ref{dekalk3}) among the three independent linear combinations
(\ref{dekalk2})-(\ref{dekalk4}) gives the weakest bounds on $\Laeff^2$
and $\La_W$, its strong anticorrelation with $S$ causes it to give in
combination with the limit on $S$ the best limit on $\De\ka$.
This is true also for the case considered next.

We notice that the careful separation of longitudinal and transversal 
structure functions allows us to put a limit on $\De\ka$ independent
of assumptions on the size of the cut-off.
This is in contrast with other methods, where an arbitrary estimate of the
size of the cut-off is made, typically of the order of a TeV.

\subsctn{The Case $\La_V^2 < 0$, $\La_W^2 > 0$}
The analysis in this case proceeds exactly analogous to the previous case.
Only here we change the notation to $\La_V^2 \rightarrow - \La_V^2$.
Following the same steps as before, we now find
\beq
\Laeff^2<1.8\mw^2\,,
\eeq
\beq
\La_W > 1.5 TeV\,,
\eeq
\beq
\La_V > 0.74 TeV\,,
\eeq
\beq
\label{bound2}
|\De\ka| = \frac {0.25}{\La_W(TeV) \La_V(TeV)} \ltap 0.08\,.
\eeq
When combining (\ref{bound1}) and (\ref{bound2}), we have in principle
to take into account that we do not know which case is realized.
Since (\ref{bound2}) is significantly more stringent than (\ref{bound1}),
the case $\La_V^2 < 0$, $\La_W^2 > 0$ with $|\De\ka|>0.26$ has negligible
probability and the bound (\ref{bound1}) gives a 95\% confidence level
overall bound.

\subsctn{Anomalous Contribution to the Photon Structure Function}
Here we relate our results to two works dealing with the changes
to the photon structure function induced by new physics.

To make contact with an earlier paper by one of the authors \cite{Bi2}
we use again the identity $\De\ka=g_1/g$.
Besides this we identify $\La$ there with $\La_V$ in the present article.
Translating the limit found there,
\beq
|\De\ka(\La/\mw)|\ltap 33
\eeq
gives
\beq
|\Laeff|\ltap 21\mw
\eeq
and we see that our bounds improve more than an order of magnitude
on this.

Measurements of the running of $\al$ can be used to put limits
on $\Laeff$.
In \cite{Ko} bounds at the 95\% confidence level on the effective scale
where new physics comes in were given as
\bea
\La_->702GeV\,,\\
\La_+>535GeV\,.
\eea
Identifying $\La_-$ or $\La_+$ with $\La_{\rm expt}$ and $\La_V$ with
$\La$ in the relation
\beq
\La_{\rm expt}=\frac{8\pi \mw^2}{e\La\De\ka}
\eeq
from \cite{Bi2} gives limits
\bea
\Laeff<6.0\mw\,,\\
\Laeff<7.9\mw\,,
\eea
which are considerably weaker than our bounds.

\subsctn{Relation to Direct Searches}
The only gauge-boson self-coupling parameter being measured directly
that can be compared to our results is $\De\ka_\ga$ in the phenomenological
Lagrange density \cite{lep2,GaGo..}
\bea
\Lag
&=&
-igc[\De g_1^Z Z^\mu(W_{\mu\nu}^-W^{+\nu}-W_{\mu\nu}^+W^{-\nu})
+\De\ka_ZW_\mu^+W_\nu^-Z^{\mu\nu}]
+igs\De\ka_\ga W_\mu^+W_\nu^-F^{\mu\nu}
\nn\\
&&
-\frac{i\la_Z}{m_W^2}Z_\mu^\nu W_\nu^{+\rho}W_\rho^{-\mu}
-\frac{i\la_\ga}{m_W^2}F_\mu^\nu W_\nu^{+\rho}W_\rho^{-\mu}\,,
\eea
where $F^{\mu\nu}$ is the electromagnetic field strength.
The relations to our triple gauge boson couplings are
\bea
g_1&=&c^2g\De g_1^Z\,,
\\
g_2&=&csg(\De\ka_Z-\De\ka_\ga)\,,
\\
g_3&=&-c^2g\De g_1^Z+c^2g\De\ka_Z+s^2g\De\ka_\ga\,,
\\
\la_Z&=&\la_\ga=0\,.
\eea
Custodial symmetry for $g'\rightarrow0$ requires $g_2=g_3=0$, leading to
\beq
\De\ka \equiv \De\ka_\ga = \De\ka_Z = c^2\De g_1^Z\,,
\eeq
and thus
\beq
\De\ka=g_1/g\,.
\eeq

Another popular set of parameters is
\bea
\al_{w\phi} &=& c^2\De g_1^Z\,,
\\
\al_w &=& \la_\ga\,,
\\
\al_{b\phi} &=& \De\ka_\ga-c^2\De g_1^Z\,,
\eea
together with the constraints
\beq
\label{constraint}
c^2\De g_1^Z = c^2\De\ka_Z+s^2\De\ka_\ga\,,
\eeq
\beq
\la_Z = \la_\ga=0\,.
\eeq
While from (\ref{constraint}) already follows $g_3=0$, the demand that
also $g_2=\la_Z=\la_\ga=0$ yields 
\beq
\al_{w\phi} = \De\ka = g_1/g\,,
\eeq
\beq
\al_w = \al_{b\phi}=0\,.
\eeq

The best available Fermilab bound combined from several Tevatron runs
is compiled by the D0 collaboration and reads \cite{fermilab}
\beq
-0.33 < \De\ka < 0.45
\eeq
at 95\% confidence level.
This bound assumes that $\De g_1^Z=0$.
As can be inferred from figure 3d in \cite{fermilab}, our assumption
that $\De\ka=c^2\De g_1^Z$ leads to a bound that is roughly twice as
stringent.
However, we note that this limit assumes a cut-off of $1.5TeV$ in the analysis.
This maybe too optimistic, as we have seen that the longitudinal cut-off
could be smaller.
If we assume that one can take $\La_V>1.5TeV$ and use the results
from $U$, $V$, $X$, we would find $\De\ka<0.13$.
Therefore the Fermilab data appear to be on the verge of being competitive now.

The best limit from CERN experiments so far is provided by the LEP2
collaboration ALEPH from combined hadronically and semileptonically
decaying $W^+W^-$ pairs and reads \cite{cern}
\beq
-0.62(0.14) < \al_{w\phi} < 0.41(0.12)
\eeq
at 95\% confidence level, where the numbers in parentheses give systematic
uncertainties.

We conclude therefore that at present the best limit on $\De\ka$ still comes
from the high precision LEP-100 data.
However LEP-200 is already competitive and should be able to improve the
limits \cite{lep2}.
The situation at Fermilab is somewhat less clear, as the limits depend on
the assumed form factors.
An analysis of the Fermilab data in terms of our cut-off propagators with
$\La_B$, $\La_W$, $\La_V$ should be useful in order to clarify the situation.
This is in particular important, in order to determine the ultimate precision
on the anomalous couplings that can be reached after the upgrade of the
Tevatron.

\subsctn{Comparison with other methods}
Finally we make a comparison with other results in the literature.

In \cite{HaIs..} the quadratic divergences are regulated by introducing 
the Higgs particle in the Lagrangian.
The anomalous couplings are in this model generated through spontaneous
symmetry breaking from higher dimension operators coupling vector boson
operators with the Higgs sector.
This regulates some of the quadratic divergences, but others still have
to be treated by other means, i.e.\ as poles in $(d-2)$ in dimensional
regularization.
This way two cut-offs appear, $m_H$ and $\La$.
This method should qualitatively give the same results as our method with
the replacements $m_H \rightarrow \La_V$ and
$\La\rightarrow\La_{B,W}$.
Unfortunately ref.\ \cite{HaIs..} calculated only the terms which are linear
in the anomalous couplings, which are less divergent, so we can only compare
the $g_1 g$ term in the S parameter.
This term is actually of the expected form.
Moreover it is found in \cite{HaIs..} that the higher divergences are physical.
The contribution to $T$ from $g_2$ found in \cite{HaIs..} is of a higher degree
in the cut-off than the contribution from $g_1$.
This supports the arguments concerning the breaking of the $SU_R(2)$
invariance.
A numerical comparison is impossible, given the fact that quantities with
different cut-off dependence were calculated. It should be interesting to
compare the results for $V,W,X$ with the scheme of \cite{HaIs..}.

In ref.\ \cite{HeVe} the quadratic divergences were regulated
by replacing poles in $(d-2)$ by $\La^2$.
This should roughly correspond with our results for $\La_W = \La_V$.
Translated in our notation ref.\ \cite{HeVe} finds
$-0.013 < \De\ka < 0.033$ for  a cut-off of $3 TeV$.
If we use our formula (62) we find $|\De\ka| < 0.028$.
So there is at least a qualitative agreement.

In \cite{DaVa} quadratic divergences are not considered, as dimensional 
regularization is used.
In the case only $g_1$ is considered it is found in our notation 
$-0.07 < \De\ka < 0.05$ for a cut-off of $2TeV$.
If we use our formula (62) we find  $|\De\ka| < 0.06$.
This agreement is accidental, as the regularization methods are quite
different. In \cite{DaVa} the logarithmically divergent terms containing
one power of the anomalous coupling are studied, whereas we consider the
more divergent terms containing two anomalous couplings.
This difference becomes clearer, when one considers the contributions from
the four-point vertices $g_4$ and $g_5$.
Both we and ref.\ \cite{DaVa} find that the corrections appear in the
combination $5g_4+2g_5$, thereby confirming the previous results from
ref.\ \cite{Bi1}.
Translated in our notation ref.\ \cite{DaVa} quotes a limit of 
$-0.15 < 5g_4+2g_5 < 0.14$, for a cut-off of $2 TeV$. 
Ignoring the logarithmic enhancement of the correction, but keeping the
quadratic part we find the stronger limit,
$ -0.066 < (5g_4+2g_5)\La^2_B (TeV) < 0.026 $.
The difference is clearly due to the different treatment of the quadratic
divergences.
As there are however more terms contributing to $T$, one should be careful in
the interpretation of this limit.

\section*{Acknowledgements}
We are grateful to T.~Takeuchi for providing us with up-to-date values
of experimental constraints on oblique parameters and to G.~Bella and
T.~Yasuda for pointing out the improved stringency of the Fermilab bound
under our assumptions.
B.K.\ thanks T.~Binoth and G.~Jikia for numerous helpful discussions.
This work was supported by the Deutsche Forschungsgemeinschaft (DFG).

\appendix

\section*{Appendix}

\sctn{Results}
Here we present our results for the vacuum polarizations.
Only the $\Pi_{XY}^g(k^2)$ are needed, since the contribution of
the $\Pi_{XY}^k(k^2)$ part is suppressed in experimentally accessible
observables by the smallness of the involved fermion masses.

When evaluating integrals, we assume that
$\xi\ll\mw^2/\La_V^2,\mz^2/\La_V^2$.
If this is not the case, the more than logarithmic divergences in
one-loop graphs are not limited to vacuum polarization corrections
for terms containing both anomalous and gauge couplings.

Tables \ref{azgraphs} and \ref{wwgraphs} show the one-loop
vacuum polarization diagrams that can be constructed from the Feynman
rules given in appendix \ref{feynmanrules}.
The integrals needed for their evaluation can be found in appendix
\ref{integrals}.
\begin{table}[htb]
\begin{center}
\mbox{\ }\vspace{-30pt}\\
\bpi(100,50)(-4,16)
\put(48,24){\circle{32}}
\put(16,8){\line(1,0){64}}
\put(48,8){\circle*{3}}
\put(40,13){$\sss+$}
\put(50,13){$\sss-$}
\put(45,45){$\scs W$}
\put(0,6){$X$}
\put(86,6){$Y$}
\epi
\hspace{5pt},
\bpi(100,50)(-4,16)
\put(48,24){\circle{32}}
\put(16,8){\line(1,0){64}}
\put(48,8){\circle*{3}}
\put(40,13){$\sss+$}
\put(50,13){$\sss-$}
\put(45,45){$\scs v$}
\put(0,6){$X$}
\put(86,6){$Y$}
\epi
\hspace{5pt},
\bpi(100,50)(-4,16)
\put(48,24){\circle{32}}
\put(16,24){\line(1,0){16}}
\put(64,24){\line(1,0){16}}
\put(32,24){\circle*{3}}
\put(36,28){$\sss+$}
\put(36,18){$\sss-$}
\put(64,24){\circle*{3}}
\put(54,28){$\sss-$}
\put(54,18){$\sss+$}
\put(0,22){$X$}
\put(86,22){$Y$}
\put(45,45){$\scs W$}
\put(45,-2){$\scs W$}
\epi
\hspace{5pt},
\bpi(100,50)(-4,16)
\put(48,24){\circle{32}}
\put(16,24){\line(1,0){16}}
\put(64,24){\line(1,0){16}}
\put(32,24){\circle*{3}}
\put(36,28){$\sss+$}
\put(36,18){$\sss-$}
\put(64,24){\circle*{3}}
\put(54,28){$\sss-$}
\put(54,18){$\sss+$}
\put(0,22){$X$}
\put(86,22){$Y$}
\put(45,45){$\scs v$}
\put(45,-2){$\scs W$}
\epi
\hspace{5pt},
\vspace{5pt}\\
\bpi(100,50)(-4,16)
\put(48,24){\circle{32}}
\put(16,24){\line(1,0){16}}
\put(64,24){\line(1,0){16}}
\put(32,24){\circle*{3}}
\put(36,28){$\sss-$}
\put(36,18){$\sss+$}
\put(64,24){\circle*{3}}
\put(54,28){$\sss+$}
\put(54,18){$\sss-$}
\put(0,22){$X$}
\put(86,22){$Y$}
\put(45,45){$\scs v$}
\put(45,-2){$\scs W$}
\epi
\hspace{5pt},
\bpi(100,50)(-4,16)
\put(48,24){\circle{32}}
\put(16,24){\line(1,0){16}}
\put(64,24){\line(1,0){16}}
\put(32,24){\circle*{3}}
\put(36,28){$\sss+$}
\put(36,18){$\sss-$}
\put(64,24){\circle*{3}}
\put(54,28){$\sss-$}
\put(54,18){$\sss+$}
\put(0,22){$X$}
\put(86,22){$Y$}
\put(45,45){$\scs v$}
\put(45,-2){$\scs v$}
\epi
\hspace{5pt},
\bpi(100,50)(-4,16)
\put(48,24){\circle{32}}
\put(16,24){\line(1,0){16}}
\put(64,24){\line(1,0){16}}
\put(32,24){\circle*{3}}
\put(36,28){$\sss+$}
\put(36,18){$\sss-$}
\put(64,24){\circle*{3}}
\put(54,28){$\sss-$}
\put(54,18){$\sss+$}
\put(45,38){$\scs<$}
\put(45,6){$\scs>$}
\put(0,22){$X$}
\put(86,22){$Y$}
\put(45,45){$\scs\eta_{\sss W}$}
\put(45,-2){$\scs\eta_{\sss W}$}
\epi
\hspace{5pt},
\bpi(100,50)(-4,16)
\put(48,24){\circle{32}}
\put(16,24){\line(1,0){16}}
\put(64,24){\line(1,0){16}}
\put(32,24){\circle*{3}}
\put(36,28){$\sss-$}
\put(36,18){$\sss+$}
\put(64,24){\circle*{3}}
\put(54,28){$\sss+$}
\put(54,18){$\sss-$}
\put(45,38){$\scs<$}
\put(45,6){$\scs>$}
\put(0,22){$X$}
\put(86,22){$Y$}
\put(45,45){$\scs\eta_{\sss W}$}
\put(45,-2){$\scs\eta_{\sss W}$}
\epi
\hspace{5pt},
\vspace{-5pt}\\
\rule[-10pt]{0pt}{70pt}
\bpi(100,50)(-4,16)
\put(48,24){\circle{32}}
\put(16,8){\line(1,0){64}}
\put(48,8){\circle*{3}}
\put(40,13){$\sss+$}
\put(50,13){$\sss-$}
\put(45,45){$\scs Z$}
\put(0,6){$X$}
\put(86,6){$Y$}
\epi
\hspace{5pt},
\bpi(100,50)(-4,16)
\put(48,24){\circle{32}}
\put(16,8){\line(1,0){64}}
\put(48,8){\circle*{3}}
\put(40,13){$\sss+$}
\put(50,13){$\sss-$}
\put(45,45){$\scs v_3$}
\put(0,6){$X$}
\put(86,6){$Y$}
\epi
\hspace{5pt}.
\rule[-10pt]{0pt}{70pt}
\end{center}
\caption{One-loop diagrams contributing to $\Pi_{XY}(k^2)$, where
$XY=AA,ZA,ZZ$.
The last two diagrams exist only for $XY=ZZ$.}
\label{azgraphs}
\end{table}
\begin{table}[htb]
\begin{center}
\mbox{\ }\vspace{-30pt}\\
\bpi(127,95)(-4,-5)
\put(48,56){\circle{32}}
\put(16,8){\line(1,0){64}}
\put(48,8){\line(0,1){32}}
\put(18,2){$\sss+$}
\put(73,2){$\sss-$}
\put(48,8){\circle*{3}}
\put(48,40){\circle*{3}}
\put(40,45){$\sss+$}
\put(50,45){$\sss-$}
\put(45,77){$\scs W$}
\put(0,6){$W$}
\put(86,6){$W$\hspace{8pt},}
\put(52,24){$\scs Z,ZA,AZ,A,v_3$}
\epi
\bpi(127,95)(-4,-5)
\put(48,56){\circle{32}}
\put(16,8){\line(1,0){64}}
\put(48,8){\line(0,1){32}}
\put(18,2){$\sss+$}
\put(73,2){$\sss-$}
\put(48,8){\circle*{3}}
\put(48,40){\circle*{3}}
\put(40,45){$\sss+$}
\put(50,45){$\sss-$}
\put(45,77){$\scs v_\pm$}
\put(0,6){$W$}
\put(86,6){$W$\hspace{8pt},}
\put(52,24){$\scs Z,ZA,AZ,A$}
\epi
\bpi(127,95)(-4,-5)
\put(48,56){\circle{32}}
\put(16,8){\line(1,0){64}}
\put(48,8){\line(0,1){32}}
\put(18,2){$\sss+$}
\put(73,2){$\sss-$}
\put(48,8){\circle*{3}}
\put(48,40){\circle*{3}}
\put(40,45){$\sss+$}
\put(50,45){$\sss-$}
\put(45,77){$\scs\eta_{\sss W}$}
\put(45,70){$\scs<$}
\put(0,6){$W$}
\put(86,6){$W$\hspace{8pt},}
\put(52,24){$\scs Z,ZA,AZ,A,v_3$}
\epi
\bpi(127,95)(-4,-5)
\put(48,56){\circle{32}}
\put(16,8){\line(1,0){64}}
\put(48,8){\line(0,1){32}}
\put(18,2){$\sss+$}
\put(73,2){$\sss-$}
\put(48,8){\circle*{3}}
\put(48,40){\circle*{3}}
\put(40,45){$\sss-$}
\put(50,45){$\sss+$}
\put(45,77){$\scs\eta_{\sss W}$}
\put(45,70){$\scs<$}
\put(0,6){$W$}
\put(86,6){$W$\hspace{8pt},}
\put(52,24){$\scs Z,ZA,AZ,A,v_3$}
\epi
\vspace{-15pt}\\
\bpi(100,50)(-4,16)
\put(48,24){\circle{32}}
\put(16,8){\line(1,0){64}}
\put(18,2){$\sss+$}
\put(73,2){$\sss-$}
\put(48,8){\circle*{3}}
\put(40,13){$\sss+$}
\put(50,13){$\sss-$}
\put(45,45){$\scs W$}
\put(0,6){$W$}
\put(86,6){$W$}
\epi
\hspace{5pt},
\bpi(100,50)(-4,16)
\put(48,24){\circle{32}}
\put(16,8){\line(1,0){64}}
\put(18,2){$\sss+$}
\put(73,2){$\sss-$}
\put(48,8){\circle*{3}}
\put(45,45){$\scs Z$}
\put(0,6){$W$}
\put(86,6){$W$}
\epi
\hspace{5pt},
\bpi(100,50)(-4,16)
\put(48,24){\circle{32}}
\put(16,8){\line(1,0){64}}
\put(18,2){$\sss+$}
\put(73,2){$\sss-$}
\put(48,8){\circle*{3}}
\put(41,45){$\scs ZA$}
\put(0,6){$W$}
\put(86,6){$W$}
\epi
\hspace{5pt},
\bpi(100,50)(-4,16)
\put(48,24){\circle{32}}
\put(16,8){\line(1,0){64}}
\put(18,2){$\sss+$}
\put(73,2){$\sss-$}
\put(48,8){\circle*{3}}
\put(45,45){$\scs A$}
\put(0,6){$W$}
\put(86,6){$W$}
\epi
\hspace{5pt},
\vspace{-10pt}\\
\rule[-10pt]{0pt}{70pt}
\bpi(100,50)(-4,16)
\put(48,24){\circle{32}}
\put(16,8){\line(1,0){64}}
\put(18,2){$\sss+$}
\put(73,2){$\sss-$}
\put(48,8){\circle*{3}}
\put(40,13){$\sss+$}
\put(50,13){$\sss-$}
\put(45,45){$\scs v_\pm$}
\put(0,6){$W$}
\put(86,6){$W$}
\epi
\hspace{5pt},
\bpi(100,50)(-4,16)
\put(48,24){\circle{32}}
\put(16,8){\line(1,0){64}}
\put(18,2){$\sss+$}
\put(73,2){$\sss-$}
\put(48,8){\circle*{3}}
\put(45,45){$\scs v_3$}
\put(0,6){$W$}
\put(86,6){$W$}
\epi
\hspace{5pt},
\bpi(100,50)(-4,16)
\put(48,24){\circle{32}}
\put(16,24){\line(1,0){16}}
\put(64,24){\line(1,0){16}}
\put(18,18){$\sss+$}
\put(73,18){$\sss-$}
\put(32,24){\circle*{3}}
\put(36,18){$\sss-$}
\put(64,24){\circle*{3}}
\put(54,18){$\sss+$}
\put(0,22){$W$}
\put(86,22){$W$}
\put(45,45){$\scs Z$}
\put(45,-2){$\scs W$}
\epi
\hspace{5pt},
\bpi(100,50)(-4,16)
\put(48,24){\circle{32}}
\put(16,24){\line(1,0){16}}
\put(64,24){\line(1,0){16}}
\put(18,18){$\sss+$}
\put(73,18){$\sss-$}
\put(32,24){\circle*{3}}
\put(36,18){$\sss-$}
\put(64,24){\circle*{3}}
\put(54,18){$\sss+$}
\put(0,22){$W$}
\put(86,22){$W$}
\put(41,45){$\scs ZA$}
\put(45,-2){$\scs W$}
\epi
\hspace{5pt},
\vspace{-10pt}\\
\rule[-10pt]{0pt}{70pt}
\bpi(100,50)(-4,16)
\put(48,24){\circle{32}}
\put(16,24){\line(1,0){16}}
\put(64,24){\line(1,0){16}}
\put(18,18){$\sss+$}
\put(73,18){$\sss-$}
\put(32,24){\circle*{3}}
\put(36,18){$\sss-$}
\put(64,24){\circle*{3}}
\put(54,18){$\sss+$}
\put(0,22){$W$}
\put(86,22){$W$}
\put(41,45){$\scs AZ$}
\put(45,-2){$\scs W$}
\epi
\hspace{5pt},
\bpi(100,50)(-4,16)
\put(48,24){\circle{32}}
\put(16,24){\line(1,0){16}}
\put(64,24){\line(1,0){16}}
\put(18,18){$\sss+$}
\put(73,18){$\sss-$}
\put(32,24){\circle*{3}}
\put(36,18){$\sss-$}
\put(64,24){\circle*{3}}
\put(54,18){$\sss+$}
\put(0,22){$W$}
\put(86,22){$W$}
\put(45,45){$\scs A$}
\put(45,-2){$\scs W$}
\epi
\hspace{5pt},
\bpi(100,50)(-4,16)
\put(48,24){\circle{32}}
\put(16,24){\line(1,0){16}}
\put(64,24){\line(1,0){16}}
\put(18,18){$\sss+$}
\put(73,18){$\sss-$}
\put(32,24){\circle*{3}}
\put(36,28){$\sss-$}
\put(64,24){\circle*{3}}
\put(54,28){$\sss+$}
\put(0,22){$W$}
\put(86,22){$W$}
\put(45,45){$\scs W$}
\put(45,-2){$\scs v_3$}
\epi
\hspace{5pt},
\bpi(100,50)(-4,16)
\put(48,24){\circle{32}}
\put(16,24){\line(1,0){16}}
\put(64,24){\line(1,0){16}}
\put(18,18){$\sss+$}
\put(73,18){$\sss-$}
\put(32,24){\circle*{3}}
\put(36,18){$\sss-$}
\put(64,24){\circle*{3}}
\put(54,18){$\sss+$}
\put(0,22){$W$}
\put(86,22){$W$}
\put(45,45){$\scs Z$}
\put(45,-2){$\scs v_\pm$}
\epi
\hspace{5pt},
\vspace{-10pt}\\
\rule[-10pt]{0pt}{70pt}
\bpi(100,50)(-4,16)
\put(48,24){\circle{32}}
\put(16,24){\line(1,0){16}}
\put(64,24){\line(1,0){16}}
\put(18,18){$\sss+$}
\put(73,18){$\sss-$}
\put(32,24){\circle*{3}}
\put(36,18){$\sss-$}
\put(64,24){\circle*{3}}
\put(54,18){$\sss+$}
\put(0,22){$W$}
\put(86,22){$W$}
\put(45,45){$\scs A$}
\put(45,-2){$\scs v_\pm$}
\epi
\hspace{5pt},
\bpi(100,50)(-4,16)
\put(48,24){\circle{32}}
\put(16,24){\line(1,0){16}}
\put(64,24){\line(1,0){16}}
\put(18,18){$\sss+$}
\put(73,18){$\sss-$}
\put(32,24){\circle*{3}}
\put(36,18){$\sss-$}
\put(64,24){\circle*{3}}
\put(54,18){$\sss+$}
\put(0,22){$W$}
\put(86,22){$W$}
\put(41,45){$\scs ZA$}
\put(45,-2){$\scs v_\pm$}
\epi
\hspace{5pt},
\bpi(100,50)(-4,16)
\put(48,24){\circle{32}}
\put(16,24){\line(1,0){16}}
\put(64,24){\line(1,0){16}}
\put(18,18){$\sss+$}
\put(73,18){$\sss-$}
\put(32,24){\circle*{3}}
\put(36,18){$\sss-$}
\put(64,24){\circle*{3}}
\put(54,18){$\sss+$}
\put(0,22){$W$}
\put(86,22){$W$}
\put(41,45){$\scs AZ$}
\put(45,-2){$\scs v_\pm$}
\epi
\hspace{5pt},
\bpi(100,50)(-4,16)
\put(48,24){\circle{32}}
\put(16,24){\line(1,0){16}}
\put(64,24){\line(1,0){16}}
\put(18,18){$\sss+$}
\put(73,18){$\sss-$}
\put(32,24){\circle*{3}}
\put(36,18){$\sss-$}
\put(64,24){\circle*{3}}
\put(54,18){$\sss+$}
\put(0,22){$W$}
\put(86,22){$W$}
\put(45,45){$\scs v_3$}
\put(45,-2){$\scs v_\pm$}
\epi
\hspace{5pt},
\vspace{-10pt}\\
\rule[-10pt]{0pt}{70pt}
\bpi(100,50)(-4,16)
\put(48,24){\circle{32}}
\put(16,24){\line(1,0){16}}
\put(64,24){\line(1,0){16}}
\put(18,18){$\sss+$}
\put(73,18){$\sss-$}
\put(32,24){\circle*{3}}
\put(36,28){$\sss-$}
\put(64,24){\circle*{3}}
\put(54,28){$\sss+$}
\put(0,22){$W$}
\put(86,22){$W$}
\put(45,38){$\scs<$}
\put(45,6){$\scs>$}
\put(45,45){$\scs\eta_{\sss W}$}
\put(45,-2){$\scs\eta_Z$}
\epi
\hspace{5pt},
\bpi(100,50)(-4,16)
\put(48,24){\circle{32}}
\put(16,24){\line(1,0){16}}
\put(64,24){\line(1,0){16}}
\put(18,18){$\sss+$}
\put(73,18){$\sss-$}
\put(32,24){\circle*{3}}
\put(36,18){$\sss-$}
\put(64,24){\circle*{3}}
\put(54,18){$\sss+$}
\put(45,38){$\scs<$}
\put(45,6){$\scs>$}
\put(0,22){$W$}
\put(86,22){$W$}
\put(45,45){$\scs\eta_Z$}
\put(45,-2){$\scs\eta_{\sss W}$}
\epi
\hspace{5pt},
\bpi(100,50)(-4,16)
\put(48,24){\circle{32}}
\put(16,24){\line(1,0){16}}
\put(64,24){\line(1,0){16}}
\put(18,18){$\sss+$}
\put(73,18){$\sss-$}
\put(32,24){\circle*{3}}
\put(36,28){$\sss-$}
\put(64,24){\circle*{3}}
\put(54,28){$\sss+$}
\put(0,22){$W$}
\put(86,22){$W$}
\put(45,38){$\scs<$}
\put(45,6){$\scs>$}
\put(45,45){$\scs\eta_{\sss W}$}
\put(45,-2){$\scs\eta_A$}
\epi
\hspace{5pt},
\bpi(100,50)(-4,16)
\put(48,24){\circle{32}}
\put(16,24){\line(1,0){16}}
\put(64,24){\line(1,0){16}}
\put(18,18){$\sss+$}
\put(73,18){$\sss-$}
\put(32,24){\circle*{3}}
\put(36,18){$\sss-$}
\put(64,24){\circle*{3}}
\put(54,18){$\sss+$}
\put(45,38){$\scs<$}
\put(45,6){$\scs>$}
\put(0,22){$W$}
\put(86,22){$W$}
\put(45,45){$\scs\eta_A$}
\put(45,-2){$\scs\eta_{\sss W}$}
\epi
\hspace{5pt}.
\end{center}
\caption{One-loop diagrams contributing to $\Pi_{WW}(k^2)$.
The tadpole graphs in the first line turn out to vanish.}
\label{wwgraphs}
\end{table}

\subsctn{$\Pi_{XY}^g(k^2)$: Quartically Divergent Terms}
\label{quartic}
The quartically divergent contributions to the vacuum polarizations
terms when all of the couplings (\ref{ano1})-(\ref{ano8}) are present
are given by
\bea
\label{piaa4}
(4\pi)^2\Pi_{AA}^g(k^2)
&=&\ts
\od(\La^2)\,,
\\
\label{piza4}
(4\pi)^2\Pi_{ZA}^g(k^2)
&=&\ts
\od(\La^2)\,,
\\
\label{pizz4}
(4\pi)^2\Pi_{ZZ}^g(k^2)
&=&\ts
g_1^2\frac{\La_V^2\La_W^2}{c^2\mw^2}\Bigg(
-\frac{3}{2\ep}-\frac{5}{4}+\frac{3}{2}
\frac{\La_V^2\ln\frac{\La_V^2}{\mubar^2}
-\La_W^2\ln\frac{\La_W^2}{\mubar^2}}
{\La_V^2-\La_W^2}\Bigg)
\nn\\
&&\ts
+g_4\frac{\La_V^4}{c^2\mw^2}
\left(\frac{2}{\ep}+\frac{5}{2}-2\ln\frac{\La_V^2}{\mubar^2}\right)
+g_5\frac{\La_V^4}{c^2\mw^2}
\left(\frac{7}{2\ep}+\frac{15}{4}
-\frac{7}{2}\ln\frac{\La_V^2}{\mubar^2}\right)
\nn\\
&&\ts
+g_6\frac{\La_V^4}{c^2\mw^2}
\left(\frac{7}{2\ep}+\frac{17}{4}
-\frac{7}{2}\ln\frac{\La_V^2}{\mubar^2}\right)
+g_7\frac{\La_V^4}{c^2\mw^2}
\left(\frac{5}{\ep}+\frac{11}{2}
-5\ln\frac{\La_V^2}{\mubar^2}\right)
\nn\\
&&\ts
+g_8\frac{\La_V^4}{c^2\mw^2}
\left(\frac{3}{\ep}+\frac{7}{2}
-3\ln\frac{\La_V^2}{\mubar^2}\right)
+\od(\La^2)\,,
\\
\label{piww4}
(4\pi)^2\Pi_{WW}^g(k^2)
&=&\ts
\left(g_1^2+g_1g_3+\frac{g_3^2}{2}\right)
\frac{\La_V^2\La_W^2}{\mw^2}\left(
-\frac{3}{2\ep}-\frac{5}{4}+\frac{3}{2}
\frac{\La_V^2\ln\frac{\La_V^2}{\mubar^2}
-\La_W^2\ln\frac{\La_W^2}{\mubar^2}}
{\La_V^2-\La_W^2}\right)
\nn\\
&&\ts
+g_2^2\frac{\La_V^2\La_B^2}{\mw^2}\left(
-\frac{3}{4\ep}-\frac{5}{8}+\frac{3}{4}
\frac{\La_V^2\ln\frac{\La_V^2}{\mubar^2}
-\La_B^2\ln\frac{\La_B^2}{\mubar^2}}
{\La_V^2-\La_B^2}\right)
\nn\\
&&\ts
+g_4\frac{\La_V^4}{\mw^2}
\left(\frac{2}{\ep}+\frac{5}{2}-2\ln\frac{\La_V^2}{\mubar^2}\right)
+g_5\frac{\La_V^4}{\mw^2}
\left(\frac{7}{2\ep}+\frac{15}{4}
-\frac{7}{2}\ln\frac{\La_V^2}{\mubar^2}\right)
\nn\\
&&\ts
+g_6\frac{\La_V^4}{\mw^2}
\left(\frac{1}{4\ep}+\frac{3}{8}-\frac{1}{4}
\ln\frac{\La_V^2}{\mubar^2}\right)
+g_7\frac{\La_V^4}{\mw^2}
\left(\frac{1}{\ep}+1-\ln\frac{\La_V^2}{\mubar^2}\right)
+\od(\La^2)\,.
\eea

\subsctn{$\Pi_{XY}^g(k^2)$ for $g_2=g_3=g_6=g_7=g_8=0$}
\label{g1g4g5}
Here we display the quartically and quadratically divergent parts
of the vacuum polarizations for the case when the anomalous couplings
preserve the custodial $SU(2)_R$ symmetry in the limit of vanishing
hypercharge coupling, i.e.\ when $g_2=g_3=g_6=g_7=g_8=0$.

Our results for the at least quadratically divergent contributions to the
$\Pi_{XY}^g(k^2)$ are
\bea
\label{piaa}
\lefteqn{(4\pi)^2\Pi_{AA}^g(k^2)}
\nn\\
&=&\ts
s^2\left\{g_1^2
\frac{\La_V^2\La_W^2}{\La_V^2-\La_W^2}\ln\frac{\La_V^2}{\La_W^2}
\left(\frac{k^2}{\mw^2}\right)
+g_1g2\La_V^2\left(\frac{1}{\ep}+1-\ln\frac{\La_V^2}{\mubar^2}\right)
\left(\frac{k^2}{\mw^2}\right)
-\frac{1}{4}g_1^2\La_V^2\left(\frac{k^2}{\mw^2}\right)^2\right\}
+\od(\La^0)\,,
\nn\\
\\
\label{piza}
\lefteqn{(4\pi)^2\Pi_{ZA}^g(k^2)}
\nn\\
&=&\ts
\frac{s}{c}\Bigg\{g_1^2
\La_V^2\La_W^2\left[
-\frac{1}{\La_V^2-\La_W^2}
+\left(\frac{\La_V^2+\La_W^2}{2(\La_V^2-\La_W^2)^2}
+\frac{s^2}{\La_V^2-\La_W^2}\right)
\ln\frac{\La_V^2}{\La_W^2}\right]\left(\frac{k^2}{\mw^2}\right)
\nn\\
&&\ts\hspace{15pt}
-g_1g\frac{3-4s^2}{2}\La_V^2
\left(\frac{1}{\ep}+1-\ln\frac{\La_V^2}{\mubar^2}\right)
\left(\frac{k^2}{\mw^2}\right)
+\frac{c^2}{4}g_1^2\La_V^2\left(\frac{k^2}{\mw^2}\right)^2
\Bigg\}
+\od(\La^0)\,,
\\
\label{pizz}
\lefteqn{(4\pi)^2\Pi_{ZZ}^g(k^2)}
\nn\\
&=&\ts
g_1^2\frac{\La_V^2\La_W^2}{c^2\mw^2}\Bigg(
-\frac{3}{2\ep}-\frac{5}{4}+\frac{3}{2}
\frac{\La_V^2\ln\frac{\La_V^2}{\mubar^2}-\La_W^2\ln\frac{\La_W^2}{\mubar^2}}
{\La_V^2-\La_W^2}\Bigg)
\nn\\
&&\ts
+g_4\frac{\La_V^4}{c^2\mw^2}
\left(\frac{2}{\ep}+\frac{5}{2}-2\ln\frac{\La_V^2}{\mubar^2}\right)
+g_5\frac{\La_V^4}{c^2\mw^2}
\left(\frac{7}{2\ep}+\frac{15}{4}
-\frac{7}{2}\ln\frac{\La_V^2}{\mubar^2}\right)
\nn\\
&&\ts
-g_1^2\frac{3}{2c^2}\La_W^2
-\frac{1}{c^2}g_1g\La_W^2\left(\frac{3}{\ep}+\frac{5}{2}
-3\ln\frac{\La_W^2}{\mubar^2}\right)
-\frac{1}{c^2}g^2\La_V^2\left(\frac{1}{\ep}+\frac{3}{4}
-\ln\frac{\La_V^2}{\mubar^2}\right)
\nn\\
&&\ts
-g_4\frac{s^2\La_B^2}{c^4}\left(
\frac{9}{2\ep}+\frac{9}{4}-\frac{9}{2}\ln\frac{\La_B^2}{\mubar^2}\right)
-g_4\frac{\La_W^2}{c^2}\left(
\frac{6}{\ep}+\frac{7}{2}-6\ln\frac{\La_W^2}{\mubar^2}\right)
\nn\\
&&\ts
-g_5\frac{s^2\La_B^2}{c^4}\left(
\frac{9}{2\ep}+\frac{9}{4}-\frac{9}{2}\ln\frac{\La_B^2}{\mubar^2}\right)
-g_5\frac{\La_W^2}{c^2}\left(
\frac{21}{2\ep}+\frac{17}{4}-\frac{21}{2}\ln\frac{\La_W^2}{\mubar^2}\right)
\nn\\
&&\ts
+\frac{1}{c^2}g_1^2\La_V^2\La_W^2\bigg[
\frac{3\La_V^2-8\La_W^2}{3(\La_V^2-\La_W^2)^2}
-\frac{2s^2}{\La_V^2-\La_W^2}
+\left(
\frac{-17\La_V^4+33\La_V^2\La_W^2-6\La_W^4}{6(\La_V^2-\La_W^2)^3}
{+}\frac{(\La_V^2+\La_W^2)s^2}{(\La_V^2-\La_W^2)^2}
{+}\frac{s^4}{\La_V^2-\La_W^2}
\right)\ln\frac{\La_V^2}{\La_W^2}\bigg]
\left(\frac{k^2}{\mw^2}\right)
\nn\\
&&\ts
+g_1g\La_V^2(1-2s^2)\left(\frac{1}{\ep}+1-\ln\frac{\La_V^2}{\mubar^2}\right)
\left(\frac{k^2}{\mw^2}\right)
\nn\\
&&\ts
-\frac{c^2}{4}g_1^2\La_V^2\left(\frac{k^2}{\mw^2}\right)^2
+\od(\La^0)\,,
\\
\label{piww}
\lefteqn{(4\pi)^2\Pi_{WW}^g(k^2)}
\nn\\
&=&\ts
g_1^2\frac{\La_V^2\La_W^2}{\mw^2}\left(
-\frac{3}{2\ep}-\frac{5}{4}+\frac{3}{2}
\frac{\La_V^2\ln\frac{\La_V^2}{\mubar^2}-\La_W^2\ln\frac{\La_W^2}{\mubar^2}}
{\La_V^2-\La_W^2}\right)
\nn\\
&&\ts
+g_4\frac{\La_V^4}{\mw^2}
\left(\frac{2}{\ep}+\frac{5}{2}-2\ln\frac{\La_V^2}{\mubar^2}\right)
+g_5\frac{\La_V^4}{\mw^2}
\left(\frac{7}{2\ep}+\frac{15}{4}
-\frac{7}{2}\ln\frac{\La_V^2}{\mubar^2}\right)
\nn\\
&&\ts
-g_1^2\La_W^2\left(\frac{3}{2}
+\frac{3s^2}{4c^2}\frac{\La_B^2}{\La_B^2-\La_W^2}
\ln\frac{\La_B^2}{\La_W^2}\right)
-g_1g\La_W^2\left(\frac{3}{\ep}+\frac{5}{2}
-3\ln\frac{\La_W^2}{\mubar^2}\right)
-g^2\La_V^2\left(\frac{1}{\ep}+\frac{3}{4}
-\ln\frac{\La_V^2}{\mubar^2}\right)
\nn\\
&&\ts
-g_4\frac{s^2\La_B^2}{c^2}\left(
\frac{3}{4\ep}+\frac{5}{8}-\frac{3}{4}\ln\frac{\La_B^2}{\mubar^2}\right)
-g_4\La_W^2\left(
\frac{6}{\ep}+\frac{7}{2}-6\ln\frac{\La_W^2}{\mubar^2}\right)
\nn\\
&&\ts
-g_5\frac{s^2\La_B^2}{c^2}\left(
\frac{3}{\ep}+1-3\ln\frac{\La_B^2}{\mubar^2}\right)
-g_5\La_W^2\left(
\frac{21}{2\ep}+\frac{17}{4}-\frac{21}{2}\ln\frac{\La_W^2}{\mubar^2}\right)
\nn\\
&&\ts
+g_1^2\left(
\frac{s^2}{2c^2}\frac{\La_V^2\La_B^2}{\La_V^2-\La_B^2}
\ln\frac{\La_V^2}{\La_B^2}
+\frac{1}{3}\La_V^2\La_W^2
\frac{3\La_V^2-8\La_W^2}{(\La_V^2-\La_W^2)^2}
-\frac{1}{6}\La_V^2\La_W^2
\frac{17\La_V^4-33\La_V^2\La_W^2+6\La_W^4}{(\La_V^2-\La_W^2)^3}
\ln\frac{\La_V^2}{\La_W^2}\right)
\left(\frac{k^2}{\mw^2}\right)
\nn\\
&&\ts
+g_1g
\La_V^2\left(\frac{1}{\ep}+1-\ln\frac{\La_V^2}{\mubar^2}\right)
\left(\frac{k^2}{\mw^2}\right)
\nn\\
&&\ts
-\frac{1}{4}g_1^2\La_V^2\left(\frac{k^2}{\mw^2}\right)^2
+\od(\La^0)\,.
\eea

\sctn{Gauge Fixing}
\label{gf}
To fix the gauge we introduce a variant of the class of $R_\xi$
gauges suitable to cancel the quadratic mixing terms between
would-be Goldstone bosons and longitudinal gauge bosons in the
presence of the higher covariant derivative terms.

Specificially, we use the gauge fixing term
\beq
\label{laggf}
\ts\Lgf=-\frac{1}{2\xi}F_{Wa}^2-\frac{1}{2\xi}F_B^2
\eeq
with
\beq
F_{Wa}=\p_\mu W_a^\mu-\half\xi gv^2(1+\La_V^{-2}\p^2)u_a
\eeq
and
\beq
F_B=\p_\mu B^\mu-\half\xi g'v^2(1+\La_V^{-2}\p^2)u_3\,,
\eeq
where the $u_a$ are defined by writing $U=\exp(iu_a\tau_a)$.
The necessary ghost terms are given by
\bea
\label{laggh}
\lefteqn{\Lgh=}
\nn\\
&&
-(\eb_{\sss Wa},\eb_{\sss B})\left(\ba{cc}
\de_{ab}\p_\mu D^\mu
{+}\xi\left(\frac{gv}{2}\right)^2
(1{+}\La_V^{-2}\p^2)(\de_{ab}{-}\ep_{abc}u_c)\;\;&
\xi\frac{gg'v^2}{4}(1{+}\La_V^{-2}\p^2)(\de_{a3}{+}\ep_{a3c}u_c)\\
&\\
\xi\frac{gg'v^2}{4}(1{+}\La_V^{-2}\p^2)(\de_{3b}{-}\ep_{3bc}u_c)&
\p^2{+}\xi\left(\frac{g'v}{2}\right)^2(1{+}\La_V^{-2}\p^2)
\ea\right)
\left(\!\!\ba{c}\et_{\sss Wb}\\\et_{\sss B}\ea\!\!\right)
\nn\\
&&
+\od(u^2)\eb\et
\eea
with
\beq
D^\mu\et_{\sss Wa}=\p^\mu\et_{\sss Wa}
+g\ep_{abc}\et_{\sss Wb}W_c^\mu\,.
\eeq
Due to the relative simplicity of our gauge fixing terms, the absence of
quadratically divergent integrals in the oblique parameters becomes
manifest only in Landau gauge, i.e.\ $\xi=0$ \cite{FaSl}.

\sctn{Feynman Rules}
\label{feynmanrules}
Since the Feynman rules in higher covariant derivative regularization
have an unfamiliar appearance, we give here all rules in our version
of $R_\xi$ gauge explicitly.

To avoid confusion with the momentum $s$ appearing in the four-vertices,
write now $\sw=\sin\Th_W$ and then also $\cw=\cos\Th_W$, $\tw=\tan\Th_W$.
Additionally to (\ref{zafields}) and (\ref{wfields}), we need the following
field redefinitions.
\beq
(\eb_{\sss Z},\eb_{\sss A})
=
(\eb_{\sss W3},\eb_{\sss B})
\left(\ba{cc}\cw&-\sw\\\sw&\cw\ea\right)\,,
\eeq
\beq
\left(\!\!\ba{c}\et_{\sss Z}\\\et_{\sss A}\ea\!\!\right)
=\left(\ba{cc}\cw&\sw\\-\sw&\cw\ea\right)
\left(\!\!\ba{c}\et_{\sss W3}\\\et_{\sss B}\ea\!\!\right)\,,
\eeq
\bea
v_\pm&=&\ts\frac{v}{\sqrt{2}}(u_1\mp iu_2)\,,\\
v_3&=&vu_3\,,
\eea
\bea
\eb_{\sss W\pm}&=&\ts\frac{1}{\sqrt{2}}(\eb_{\sss W1}\mp i\eb_{\sss W2})\,,
\\
\et_{\sss W\pm}&=&\ts\frac{1}{\sqrt{2}}(\et_{\sss W1}\mp i\et_{\sss W2})\,.
\eea
Define also
\beq\ts
\mw\equiv\frac{gv}{2}\,,\;\;\;\;\;\;
\mz\equiv\frac{gv}{2\cw}\,.
\eeq
The quadratic part of the Lagrangian extracted from (\ref{lagew}),
(\ref{laghctr}), (\ref{laghclg}), (\ref{laggf}), (\ref{laggh}) reads in
terms of the redefined fields
\bea
\label{lquad}
\Lag_2
&=&\ts
W_\mu^+
\left\{\left[\La_W^{-2}(\p^2)^2
{+}\left(1{+}\La_V^{-2}\mw^2\right)\p^2{+}\mw^2\right]
(g^{\mu\nu}{-}\frac{\p^\mu\p^\nu}{\p^2})
{+}\left[\left(\frac{1}{\xi}{+}\La_V^{-2}\mw^2\right)\p^2{+}\mw^2\right]
\frac{\p^\mu\p^\nu}{\p^2}\right\}W_\nu^-
\nn\\
&&\ts
+\half(Z_\mu,A_\mu)\left[{\cal D}_{ZA}^{tr}
(g^{\mu\nu}{-}\frac{\p^\mu\p^\nu}{\p^2})
+{\cal D}_{ZA}^{lg}\frac{\p^\mu\p^\nu}{\p^2}\right]
\left(\!\!\ba{c}Z_\nu\\A_\nu\ea\!\!\right)
\nn\\
&&\ts
-v_+\left[(1+\La_V^{-2}\p^2)\p^2+\xi
\mw^2(1+\La_V^{-2}\p^2)^2\right]v_-
\nn\\
&&\ts
-\half v_3\left[(1+\La_V^{-2}\p^2)\p^2
+\xi\mz^2(1+\La_V^{-2}\p^2)^2\right]v_3
\nn\\
&&\ts
-\eb_{\sss W+}\left[\left(1+\xi\La_V^{-2}\mw^2\right)\p^2
+\xi \mw^2\right]\et_{\sss W-}
-\eb_{\sss W-}\left[\left(1+\xi\La_V^{-2}\mw^2\right)\p^2
+\xi \mw^2\right]\et_{\sss W+}
\nn\\
&&\ts
-\eb_Z\left[\left(1+\xi\La_V^{-2}\mz^2\right)\p^2
+\xi \mz^2\right]\et_Z
-\eb_A\p^2\et_A
\eea
with
\beq\ts
{\cal D}_{ZA}^{tr}=
\left(\ba{cc}
\left(\frac{\cw^2}{\La_W^2}{+}\frac{\sw^2}{\La_B^2}\right)(\p^2)^2
{+}\left(1{+}\frac{\mz^2}{\La_V^2}\right)\p^2{+}\mz^2
\,&\,\left(\frac{1}{\La_B^2}{-}\frac{1}{\La_W^2}\right)\sw\cw(\p^2)^2\\
\left(\frac{1}{\La_B^2}{-}\frac{1}{\La_W^2}\right)\sw\cw(\p^2)^2&
\left(\frac{\sw^2}{\La_W^2}{+}\frac{\cw^2}{\La_B^2}\right)(\p^2)^2{+}\p^2
\ea\right)
\eeq
and
\beq
{\cal D}_{ZA}^{lg}=
\left(\ba{cc}\left(\frac{1}{\xi}+\La_V^{-2}\mz^2\right)\p^2+\mz^2&0\\
0&\frac{1}{\xi}\p^2\ea\right)\,.
\eeq

\subsctn{Propagators}
Some of the propagators have an unusual form caused by the higher
covariant derivative terms.
However, they can be decomposed into combinations of standard propagator
terms with modified masses and normalization factors as indicated below.

\bea
\De_{\mu\nu}^W(k)
&=&
-i\left[\frac{-\La_W^2}{(k^2-m_{\sss W<}^2)(k^2-m_{\sss W>}^2)}
(g_{\mu\nu}-k_\mu k_\nu/k^2)
+\frac{Z_W^{lg}\xi k_\mu k_\nu/k^2}{k^2-m_{\sss Wlg}^2}\right]
\nn\\
&=&
-i\left[
Z_W^{tr}\left(\frac{1}{k^2-m_{\sss W<}^2}-\frac{1}{k^2-m_{\sss W>}^2}\right)
(g_{\mu\nu}-k_\mu k_\nu/k^2)
+\frac{Z_W^{lg}\xi k_\mu k_\nu/k^2}{k^2-m_{\sss Wlg}^2}\right]\,,
\\
\De_{\mu\nu}^Z(k)
&=&
-i\left[\frac{\La_W^2\La_B^2-\La_Z^2k^2}
{(k^2-m_{\sss Z<}^2)(k^2-m_{\sss Z>}^2)(k^2-m_{\sss A>}^2)}
(g_{\mu\nu}-k_\mu k_\nu/k^2)
+\frac{Z_Z^{lg}\xi k_\mu k_\nu/k^2}{k^2-m_{\sss Zlg}^2}\right]
\nn\\
&=&
-i\left[\left(\frac{Z_{Z<}^{ZZ}}{k^2{-}m_{\sss Z<}^2}
{+}\frac{Z_{Z>}^{ZZ}}{k^2{-}m_{\sss Z>}^2}
{+}\frac{Z_{A>}^{ZZ}}{k^2{-}m_{\sss A>}^2}\right)
(g_{\mu\nu}{-}k_\mu k_\nu/k^2)
{+}\frac{Z_Z^{lg}\xi k_\mu k_\nu/k^2}{k^2{-}m_{\sss Zlg}^2}\right]\,,
\\
\De_{\mu\nu}^{ZA}(k)=\De_{\mu\nu}^{AZ}(k)
&=&
-i\frac{(\La_W^2-\La_B^2)k^2\sw\cw}
{(k^2-m_{\sss Z<}^2)(k^2-m_{\sss Z>}^2)(k^2-m_{\sss A>}^2)}
(g_{\mu\nu}-k_\mu k_\nu/k^2)
\nn\\
&=&
-i\left(\frac{Z_{Z<}^{ZA}}{k^2-m_{\sss Z<}^2}
+\frac{Z_{Z>}^{ZA}}{k^2-m_{\sss Z>}^2}
+\frac{Z_{A>}^{ZA}}{k^2-m_{\sss A>}^2}\right)
(g_{\mu\nu}-k_\mu k_\nu/k^2)\,,
\\
\De_{\mu\nu}^A(k)
&=&
-i\left[\frac{\La_W^2\La_B^2\left[\left(1{+}\La_V^{-2}\mz^2\right)k^2
{-}\mz^2\right]{-}\La_A^2(k^2)^2}
{k^2(k^2-m_{\sss Z<}^2)(k^2-m_{\sss Z>}^2)(k^2-m_{\sss A>}^2)}
(g_{\mu\nu}{-}k_\mu k_\nu/k^2)
+\frac{\xi k_\mu k_\nu/k^2}{k^2}\right]
\nn\\
&=&
-i\left[\left(\frac{1}{k^2}{+}\frac{Z_{Z<}^{AA}}{k^2{-}m_{\sss Z<}^2}
{+}\frac{Z_{Z>}^{AA}}{k^2{-}m_{\sss Z>}^2}
{+}\frac{Z_{A>}^{AA}}{k^2{-}m_{\sss A>}^2}\right)
(g_{\mu\nu}{-}k_\mu k_\nu/k^2)
{+}\frac{\xi k_\mu k_\nu/k^2}{k^2}\right]\,,
\\
\De^{v_\pm}(k^2)
&=&
i\frac{-Z_W^{lg}\La_V^2}{(k^2-m_{\sss Wlg}^2)(k^2-\La_V^2)}
=i\left(\frac{1}{k^2-m_{\sss Wlg}^2}-\frac{1}{k^2-\La_V^2}\right)\,,
\\
\De^{v_3}(k^2)
&=&
i\frac{-Z_Z^{lg}\La_V^2}{(k^2-m_{\sss Zlg}^2)(k^2-\La_V^2)}
=i\left(\frac{1}{k^2-m_{\sss Zlg}^2}-\frac{1}{k^2-\La_V^2}\right)\,,
\\
\De^{\et_W}(k^2)
&=&
\frac{i Z_W^{lg}}{k^2-m_{\sss Wlg}^2}\,,
\\
\De^{\et_Z}(k^2)
&=&
\frac{i Z_Z^{lg}}{k^2-m_{\sss Zlg}^2}\,,
\\
\De^{\et_A}(k^2)
&=&
\frac{i}{k^2}
\eea
with
\bea
Z_W^{tr}
&=&
\frac{\La_W^2}{m_{\sss W>}^2-m_{\sss W<}^2}
=
\frac{1}{\sqrt{\left(1+\La_V^{-2}\mw^2\right)^2-4\La_W^{-2}\mw^2}}
=
1+\od(\La^{-2})\,,
\\
Z_W^{lg}
&=&
\frac{1}{1+\xi\La_V^{-2}\mw^2}
=
1+\od(\La^{-2})\,,
\\
Z_{Z<}^{ZZ}
&=&
\frac{\La_B^2\La_W^2-\La_Z^2m_{\sss Z<}^2}
{(m_{\sss Z<}^2-m_{\sss Z>}^2)(m_{\sss Z<}^2-m_{\sss A>}^2)}
=
1+\od(\La^{-2})\,,
\\
Z_{Z>}^{ZZ}
&=&
\frac{\La_B^2\La_W^2-\La_Z^2m_{\sss Z>}^2}
{(m_{\sss Z>}^2-m_{\sss Z<}^2)(m_{\sss Z>}^2-m_{\sss A>}^2)}
=
-\cwcw+\od(\La^{-2})\,,
\\
Z_{A>}^{ZZ}
&=&
\frac{\La_B^2\La_W^2-\La_Z^2m_{\sss A>}^2}
{(m_{\sss A>}^2-m_{\sss Z<}^2)(m_{\sss A>}^2-m_{\sss Z>}^2)}
=
-\swsw+\od(\La^{-2})\,,
\\
Z_{Z<}^{ZA}
&=&
\frac{(\La_W^2-\La_B^2)m_{\sss Z<}^2\sw\cw}
{(m_{\sss Z<}^2-m_{\sss Z>}^2)(m_{\sss Z<}^2-m_{\sss A>}^2)}
=
\od(\La^{-2})\,,
\\
Z_{Z>}^{ZA}
&=&
\frac{(\La_W^2-\La_B^2)m_{\sss Z>}^2\sw\cw}
{(m_{\sss Z>}^2-m_{\sss Z<}^2)(m_{\sss Z>}^2-m_{\sss A>}^2)}
=
\sw\cw+\od(\La^{-2})\,,
\\
Z_{A>}^{ZA}
&=&
\frac{(\La_W^2-\La_B^2)m_{\sss A>}^2\sw\cw}
{(m_{\sss A>}^2-m_{\sss Z<}^2)(m_{\sss A>}^2-m_{\sss Z>}^2)}
=
-\sw\cw+\od(\La^{-2})\,,
\\
Z_{Z<}^{AA}
&=&
\frac{\La_W^2\La_B^2
-m_{\sss Z>}^2m_{\sss A>}^2(1-\La_V^{-2}m_{\sss Z<}^2)
-\La_A^2m_{\sss Z<}^2}
{(m_{\sss Z<}^2-m_{\sss Z>}^2)(m_{\sss Z<}^2-m_{\sss A>}^2)}
=
\od(\La^{-4})\,,
\\
Z_{Z>}^{AA}
&=&
\frac{\La_W^2\La_B^2
-m_{\sss Z<}^2m_{\sss A>}^2(1-\La_V^{-2}m_{\sss Z>}^2)
-\La_A^2m_{\sss Z>}^2}
{(m_{\sss Z>}^2-m_{\sss Z<}^2)(m_{\sss Z>}^2-m_{\sss A>}^2)}
=
-\swsw+\od(\La^{-2})\,,
\\
Z_{A>}^{AA}
&=&
\frac{\La_W^2\La_B^2
-m_{\sss Z<}^2m_{\sss Z>}^2(1-\La_V^{-2}m_{\sss A>}^2)
-\La_A^2m_{\sss A>}^2}
{(m_{\sss A>}^2-m_{\sss Z<}^2)(m_{\sss A>}^2-m_{\sss Z>}^2)}
=
-\cwcw+\od(\La^{-2})\,,
\\
Z_Z^{lg}
&=&
\frac{1}{1+\xi\La_V^{-2}\mz^2}
=
1+\od(\La^{-2})\,,
\\
m_{W_<^>}^2
&=&
\half\La_W^2\left[\left(1+\La_V^{-2}\mw^2\right)
\pm\sqrt{\left(1+\La_V^{-2}\mw^2\right)^2-4\La_W^{-2}\mw^2}
\right]
=\left\{\ba{l}\La_W^2\\\mw^2\ea\right\}
\times\left(1+\od(\La^{-2})\right)\,,
\nn\\ \\
m_{\sss Wlg}^2
&=&
\frac{\xi \mw^2}{1+\xi\La_V^{-2}\mw^2}\,,
\\
m_{\sss Zlg}^2
&=&
\frac{\xi \mz^2}{1+\xi\La_V^{-2}\mz^2}\,,
\\
\La_Z^2&=&\La_W^2\cwcw+\La_B^2\swsw\,,
\\
\La_A^2&=&\La_W^2\swsw+\La_B^2\cwcw\,,
\eea
and where $m_{\sss Z<}^2$, $m_{\sss Z>}^2$, $m_{\sss A>}^2$ are
determined by
\bea
\label{rxidens}
\lefteqn{(k^2-m_{\sss Z<}^2)(k^2-m_{\sss Z>}^2)(k^2-m_{\sss A>}^2)}
\nn\\
&=&
(k^2)^3
-\left[\La_W^2+\La_B^2+\La_V^{-2}\La_Z^2\mz^2\right](k^2)^2
+\left[\La_W^2\La_B^2\left(1+\La_V^{-2}\mz^2\right)
+\La_Z^2\mz^2\right]k^2-\La_W^2\La_B^2\mz^2
\nn\\
\eea
i.e.\
\bea
\La_W^2+\La_B^2+\La_V^{-2}\La_Z^2\mz^2
&=&
m_{\sss Z<}^2+m_{\sss Z>}^2+m_{\sss A>}^2\,,
\\
\La_W^2\La_B^2+\La_Z^2\mz^2+\La_V^{-2}\La_W^2\La_B^2\mz^2
&=&
m_{\sss Z<}^2m_{\sss Z>}^2
+m_{\sss Z<}^2m_{\sss A>}^2+m_{\sss Z>}^2m_{\sss A>}^2\,,
\\
\La_W^2\La_B^2\mz^2
&=&
m_{\sss Z<}^2m_{\sss Z>}^2m_{\sss A>}^2
\eea
with
\bea
m_{\sss Z<}^2
&=&
\mz^2\left(1+\od(\La^{-2})\right)\,,
\\
m_{\sss Z>}^2
&=&
\La_W^2\left(1+\od(\La^{-2})\right)\,,
\\
m_{\sss A>}^2
&=&
\La_B^2\left(1+\od(\La^{-2})\right)\,.
\eea

The masses and renormalization constants have to be evaluated to higher
order than explicitly given here.

\subsctn{Vertices}
All momenta are outgoing.
Only vertices needed for one-loop gauge propagator corrections are displayed.

\subsubsctn{Four-Vertices}
\bea
\bpi(80,40)(-20,10)
\multiput(16,18)(4,4){4}{\oval(4,4)[br]}
\multiput(20,18)(4,4){4}{\oval(4,4)[tl]}
\multiput(0,30)(4,-4){4}{\oval(4,4)[tr]}
\multiput(4,30)(4,-4){4}{\oval(4,4)[bl]}
\multiput(0,2)(4,4){4}{\oval(4,4)[br]}
\multiput(4,2)(4,4){4}{\oval(4,4)[tl]}
\multiput(16,14)(4,-4){4}{\oval(4,4)[tr]}
\multiput(20,14)(4,-4){4}{\oval(4,4)[bl]}
\put(16,16){\circle*{3}}
\put(-30,40){$W^+,p,\al$}
\put(22,40){$W^+,q,\be$}
\put(-30,-12){$W^-,r,\ga$}
\put(22,-12){$W^-,s,\de$}
\epi
&=&\ts
i\Big\{
g_{\al\be}g_{\ga\de}\Big[2(2gg_1+2gg_3+g_4)
\nn\\
&&\ts\hspace{45pt}
+g^2\left(2+4\La_V^{-2}\mw^2
+\La_W^{-2}[4(p\cdot q)+(p+q)\cdot(r+s)
+4(r\cdot s)]\right)\Big]
\nn\\
&&\ts\hspace{10pt}
-g_{\al\ga}g_{\be\de}\Big[(2gg_1+2gg_3-g_4-2g_5)
\nn\\
&&\ts\hspace{55pt}
+g^2\left(1+2\La_V^{-2}\mw^2
+\La_W^{-2}[(p\cdot q)+2(p\cdot r)+2(q\cdot s)+(r\cdot s)]\right)\Big]
\nn\\
&&\ts\hspace{10pt}
-g_{\al\de}g_{\be\ga}\Big[(2gg_1+2gg_3-g_4-2g_5)
\nn\\
&&\ts\hspace{55pt}
+g^2\left(1+2\La_V^{-2}\mw^2
+\La_W^{-2}[(p\cdot q)+2(p\cdot s)+2(q\cdot r)+(r\cdot s)]\right)\Big]
\nn\\
&&\ts\hspace{10pt}
-g^2\La_W^{-2}\Big[\hspace{10pt}
g_{\al\be}(2(p-q)_\ga(p-q)_\de-(p+q)_\ga r_\de-s_\ga(p+q)_\de+2s_\ga r_\de)
\nn\\
&&\ts\hspace{55pt}
{+}g_{\ga\de}(2(r-s)_\al(r-s)_\be-(r+s)_\al p_\be-q_\al(r+s)_\be+2p_\be q_\al)
\nn\\
&&\ts\hspace{55pt}
+g_{\al\ga}(2p_\be r_\de-p_\be p_\de-r_\be r_\de+p_\be q_\de-s_\be(q-r)_\de)
\nn\\
&&\ts\hspace{55pt}
+g_{\al\de}(2p_\be s_\ga-p_\be p_\ga-s_\be s_\ga+p_\be q_\ga-r_\be(q-s)_\ga)
\nn\\
&&\ts\hspace{55pt}
+g_{\be\ga}(2q_\al r_\de-q_\al q_\de-r_\al r_\de+q_\al p_\de-s_\al(p-r)_\de)
\nn\\
&&\ts\hspace{55pt}
+g_{\be\de}(2q_\al s_\ga-q_\al q_\ga-s_\al s_\ga+q_\al p_\ga-r_\al(p-s)_\ga)
\Big]\Big\}
\\
\bpi(80,40)(-20,10)
\multiput(16,18)(4,4){4}{\oval(4,4)[br]}
\multiput(20,18)(4,4){4}{\oval(4,4)[tl]}
\multiput(0,30)(4,-4){4}{\oval(4,4)[tr]}
\multiput(4,30)(4,-4){4}{\oval(4,4)[bl]}
\multiput(0,2)(4,4){4}{\oval(4,4)[br]}
\multiput(4,2)(4,4){4}{\oval(4,4)[tl]}
\multiput(16,14)(4,-4){4}{\oval(4,4)[tr]}
\multiput(20,14)(4,-4){4}{\oval(4,4)[bl]}
\put(16,16){\circle*{3}}
\put(-30,40){$W^+,p,\al$}
\put(22,40){$W^-,q,\be$}
\put(-18,-12){$Z,r,\ga$}
\put(22,-12){$Z,s,\de$}
\epi
&=&
-i\Big\{
g_{\al\be}g_{\ga\de}\Big[4gg_1-2(g_5+g_7)\oocwcw
\nn\\
&&\ts\hspace{55pt}
+g^2\Big(2\cwcw+2\La_V^{-2}\mw^2(\cwcw+\oocwcw)
\nn\\
&&\ts\hspace{80pt}
+\La_W^{-2}\cwcw\left(4(p\cdot q)+(p+q)\cdot(r+s)
+4(r\cdot s)\right)\Big)\Big]
\nn\\
&&\ts\hspace{15pt}
-g_{\al\ga}g_{\be\de}\Big[2gg_1+(g_4+g_6)\oocwcw
\nn\\
&&\ts\hspace{60pt}
+g^2\Big(\cwcw+2\La_V^{-2}\mw^2
+\La_W^{-2}\cwcw((p\cdot q)+2(p\cdot r)+2(q\cdot s)+(r\cdot s))\Big)\Big]
\nn\\
&&\ts\hspace{15pt}
-g_{\al\de}g_{\be\ga}\Big[2gg_1+(g_4+g_6)\oocwcw
\nn\\
&&\ts\hspace{60pt}
+g^2\Big(\cwcw+2\La_V^{-2}\mw^2
+\La_W^{-2}\cwcw((p\cdot q)+2(p\cdot s)+2(q\cdot r)+(r\cdot s))\Big)\Big]
\nn\\
&&\ts\hspace{15pt}
-g^2\La_W^{-2}\cwcw\Big[\hspace{10pt}
g_{\al\be}(2(p-q)_\ga(p-q)_\de-(p+q)_\ga r_\de-s_\ga(p+q)_\de+2s_\ga r_\de)
\nn\\
&&\ts\hspace{70pt}
+g_{\ga\de}(2(r-s)_\al(r-s)_\be-(r+s)_\al p_\be-q_\al(r+s)_\be+2p_\be q_\al)
\nn\\
&&\ts\hspace{70pt}
+g_{\al\ga}(2p_\be r_\de-p_\be p_\de-r_\be r_\de+p_\be q_\de-s_\be(q-r)_\de)
\nn\\
&&\ts\hspace{70pt}
+g_{\al\de}(2p_\be s_\ga-p_\be p_\ga-s_\be s_\ga+p_\be q_\ga-r_\be(q-s)_\ga)
\nn\\
&&\ts\hspace{70pt}
+g_{\be\ga}(2q_\al r_\de-q_\al q_\de-r_\al r_\de+q_\al p_\de-s_\al(p-r)_\de)
\nn\\
&&\ts\hspace{70pt}
+g_{\be\de}(2q_\al s_\ga-q_\al q_\ga-s_\al s_\ga+q_\al p_\ga-r_\al(p-s)_\ga)
\Big]\Big\}
\\
\bpi(80,40)(-20,10)
\multiput(16,18)(4,4){4}{\oval(4,4)[br]}
\multiput(20,18)(4,4){4}{\oval(4,4)[tl]}
\multiput(0,30)(4,-4){4}{\oval(4,4)[tr]}
\multiput(4,30)(4,-4){4}{\oval(4,4)[bl]}
\multiput(0,2)(4,4){4}{\oval(4,4)[br]}
\multiput(4,2)(4,4){4}{\oval(4,4)[tl]}
\multiput(16,14)(4,-4){4}{\oval(4,4)[tr]}
\multiput(20,14)(4,-4){4}{\oval(4,4)[bl]}
\put(16,16){\circle*{3}}
\put(-30,40){$W^+,p,\al$}
\put(22,40){$W^-,q,\be$}
\put(-18,-12){$Z,r,\ga$}
\put(22,-12){$A,s,\de$}
\epi
&=&
i\tw\Big\{
g_{\al\be}g_{\ga\de}\Big[2gg_1
+g^2\cwcw\Big(2+2\La_V^{-2}\mw^2
+\La_W^{-2}[4(p\cdot q)+(p{+}q)\cdot(r{+}s)+4(r\cdot s)]\Big]
\nn\\
&&\ts\hspace{15pt}
-g_{\al\ga}g_{\be\de}\Big[gg_1
+g^2\Big(\cwcw+\La_V^{-2}\mw^2
+\La_W^{-2}\cwcwcwcw[(p\cdot q){+}2(p\cdot r){+}2(q\cdot s){+}(r\cdot s)]
\Big)\Big]
\nn\\
&&\ts\hspace{15pt}
-g_{\al\de}g_{\be\ga}\Big[gg_1+g^2\Big(\cwcw+\La_V^{-2}\mw^2
+\La_W^{-2}\cwcwcwcw[(p\cdot q){+}2(p\cdot s){+}2(q\cdot r){+}(r\cdot s)]
\Big)\Big]
\nn\\
&&\ts\hspace{15pt}
-g^2\La_W^{-2}\cwcw\Big[\hspace{10pt}
g_{\al\be}(2(p{-}q)_\ga(p{-}q)_\de{-}(p{+}q)_\ga r_\de
{-}s_\ga(p+q)_\de{+}2s_\ga r_\de)
\nn\\
&&\ts\hspace{70pt}
+g_{\ga\de}(2(r{-}s)_\al(r{-}s)_\be{-}(r{+}s)_\al p_\be
{-}q_\al(r{+}s)_\be{+}2p_\be q_\al)
\nn\\
&&\ts\hspace{70pt}
+g_{\al\ga}(2p_\be r_\de-p_\be p_\de-r_\be r_\de+p_\be q_\de-s_\be(q-r)_\de)
\nn\\
&&\ts\hspace{70pt}
+g_{\al\de}(2p_\be s_\ga-p_\be p_\ga-s_\be s_\ga+p_\be q_\ga-r_\be(q-s)_\ga)
\nn\\
&&\ts\hspace{70pt}
+g_{\be\ga}(2q_\al r_\de-q_\al q_\de-r_\al r_\de+q_\al p_\de-s_\al(p-r)_\de)
\nn\\
&&\ts\hspace{70pt}
+g_{\be\de}(2q_\al s_\ga-q_\al q_\ga-s_\al s_\ga+q_\al p_\ga-r_\al(p-s)_\ga)
\Big]\Big\}
\nn\\&&\\
\bpi(80,40)(-20,10)
\multiput(16,18)(4,4){4}{\oval(4,4)[br]}
\multiput(20,18)(4,4){4}{\oval(4,4)[tl]}
\multiput(0,30)(4,-4){4}{\oval(4,4)[tr]}
\multiput(4,30)(4,-4){4}{\oval(4,4)[bl]}
\multiput(0,2)(4,4){4}{\oval(4,4)[br]}
\multiput(4,2)(4,4){4}{\oval(4,4)[tl]}
\multiput(16,14)(4,-4){4}{\oval(4,4)[tr]}
\multiput(20,14)(4,-4){4}{\oval(4,4)[bl]}
\put(16,16){\circle*{3}}
\put(-30,40){$W^+,p,\al$}
\put(22,40){$W^-,q,\be$}
\put(-18,-12){$A,r,\ga$}
\put(22,-12){$A,s,\de$}
\epi
&=&
-ig^2\swsw\Big\{\hspace{9pt}
g_{\al\be}g_{\ga\de}\Big(2+2\La_V^{-2}\mw^2
+\La_W^{-2}[4(p\cdot q)+(p{+}q)\cdot(r{+}s)+4(r\cdot s)]\Big)
\nn\\
&&\ts\hspace{40pt}
-g_{\al\ga}g_{\be\de}\Big(1
+\La_W^{-2}\cwcw[(p\cdot q)+2(p\cdot r)+2(q\cdot s)+(r\cdot s)]\Big)
\nn\\
&&\ts\hspace{40pt}
-g_{\al\de}g_{\be\ga}\Big(1
+\La_W^{-2}\cwcw[(p\cdot q)+2(p\cdot s)+2(q\cdot r)+(r\cdot s)]\Big)
\nn\\
&&\ts\hspace{40pt}
-\La_W^{-2}\Big[\hspace{10pt}
g_{\al\be}(2(p{-}q)_\ga(p{-}q)_\de-(p{+}q)_\ga r_\de
-s_\ga(p{+}q)_\de+2s_\ga r_\de)
\nn\\
&&\ts\hspace{75pt}
+g_{\ga\de}(2(r{-}s)_\al(r{-}s)_\be-(r{+}s)_\al p_\be
-q_\al(r{+}s)_\be+2p_\be q_\al)
\nn\\
&&\ts\hspace{75pt}
+g_{\al\ga}(2p_\be r_\de-p_\be p_\de-r_\be r_\de+p_\be q_\de-s_\be(q-r)_\de)
\nn\\
&&\ts\hspace{75pt}
+g_{\al\de}(2p_\be s_\ga-p_\be p_\ga-s_\be s_\ga+p_\be q_\ga-r_\be(q-s)_\ga)
\nn\\
&&\ts\hspace{75pt}
+g_{\be\ga}(2q_\al r_\de-q_\al q_\de-r_\al r_\de+q_\al p_\de-s_\al(p-r)_\de)
\nn\\
&&\ts\hspace{75pt}
+g_{\be\de}(2q_\al s_\ga-q_\al q_\ga-s_\al s_\ga+q_\al p_\ga-r_\al(p-s)_\ga)
\Big]\Big\}
\\
\bpi(80,40)(-20,10)
\multiput(16,18)(4,4){4}{\oval(4,4)[br]}
\multiput(20,18)(4,4){4}{\oval(4,4)[tl]}
\multiput(0,30)(4,-4){4}{\oval(4,4)[tr]}
\multiput(4,30)(4,-4){4}{\oval(4,4)[bl]}
\multiput(0,2)(4,4){4}{\oval(4,4)[br]}
\multiput(4,2)(4,4){4}{\oval(4,4)[tl]}
\multiput(16,14)(4,-4){4}{\oval(4,4)[tr]}
\multiput(20,14)(4,-4){4}{\oval(4,4)[bl]}
\put(16,16){\circle*{3}}
\put(-18,40){$Z,p,\al$}
\put(22,40){$Z,q,\be$}
\put(-18,-12){$Z,r,\ga$}
\put(22,-12){$Z,s,\de$}
\epi
&=&
2i\cw^{-4}(g_4+g_5+2g_6+2g_7+2g_8)
(g_{\al\be}g_{\ga\de}+g_{\al\ga}g_{\be\de}+g_{\al\de}g_{\be\ga})
\\&&\nn\\&&\nn\\
\bpi(80,40)(-20,10)
\multiput(0,2)(4,4){4}{\oval(4,4)[br]}
\multiput(4,2)(4,4){4}{\oval(4,4)[tl]}
\multiput(16,14)(4,-4){4}{\oval(4,4)[tr]}
\multiput(20,14)(4,-4){4}{\oval(4,4)[bl]}
\put(0,32){\line(1,-1){16}}
\put(16,16){\line(1,1){16}}
\put(16,16){\circle*{3}}
\put(-20,40){$v_+,r$}
\put(25,40){$v_-,s$}
\put(-30,-12){$W^+,p,\al$}
\put(22,-12){$W^-,q,\be$}
\epi
&=&
-\ihalf \Big\{g_{\al\be}\Big[
\mw^{-2}\Big(gg_1(p{-}q)\cdot(r{-}s){+}2gg_3(p{\cdot}r{+}q{\cdot}s)
{+}2(g_4{+}2g_5)(r{\cdot}s)\Big)
{-}2g^2\La_V^{-2}(r{\cdot}s)\Big]
\nn\\
&&\hspace{20pt}
+(gg_1\mw^{-2}+g^2\La_V^{-2})
[2(r_\al s_\be-s_\al r_\be)+q_\al(r-s)_\be-(r-s)_\al p_\be]
\nn\\
&&\hspace{20pt}
+2gg_3\mw^{-2}(r_\al s_\be{-}s_\al r_\be{-}r_\al p_\be{-}q_\al s_\be)
+2g_4(2r_\al s_\be{+}s_\al r_\be)+g_5 4s_\al r_\be\Big\}
\\&&\nn\\
\bpi(80,40)(-20,10)
\multiput(0,2)(4,4){4}{\oval(4,4)[br]}
\multiput(4,2)(4,4){4}{\oval(4,4)[tl]}
\multiput(16,14)(4,-4){4}{\oval(4,4)[tr]}
\multiput(20,14)(4,-4){4}{\oval(4,4)[bl]}
\put(0,32){\line(1,-1){16}}
\put(16,16){\line(1,1){16}}
\put(16,16){\circle*{3}}
\put(-20,40){$v_3,r$}
\put(25,40){$v_3,s$}
\put(-30,-12){$W^+,p,\al$}
\put(22,-12){$W^-,q,\be$}
\epi
&=&
2i\left[g^2\La_V^{-2}-(g_5+g_7)\right](r\cdot s)g_{\al\be}
-i\mw^{-2}(g_4+g_6)(r_\al s_\be+s_\al r_\be)
\\&&\nn\\&&\nn\\
\bpi(80,40)(-20,10)
\multiput(0,2)(4,4){4}{\oval(4,4)[br]}
\multiput(4,2)(4,4){4}{\oval(4,4)[tl]}
\multiput(16,14)(4,-4){4}{\oval(4,4)[tr]}
\multiput(20,14)(4,-4){4}{\oval(4,4)[bl]}
\put(0,32){\line(1,-1){16}}
\put(16,16){\line(1,1){16}}
\put(16,16){\circle*{3}}
\put(-20,40){$v_+,r$}
\put(25,40){$v_-,s$}
\put(-22,-12){$Z,p,\al$}
\put(27,-12){$Z,q,\be$}
\epi
&=&
-i\Big\{g_{\al\be}\Big[2g^2\swsw\Big(1+\La_V^{-2}(p\cdot q)\Big)
-2\oocwcw\Big(g^2\La_V^{-2}(\cwcwcwcw+\swswswsw)
-(g_5+g_7)\mw^{-2}\Big)(r\cdot s)
\nn\\
&&\hspace{40pt}
-\Big((gg_1\swsw-gg_2\sw\cw-gg_3\cwcw)\mw^{-2}
+g^2\La_V^{-2}\swsw\Big)
(p+q)^2\Big]
\nn\\
&&\hspace{20pt}
-2g^2\La_V^{-2}\swsw(r-s)_\al(r-s)_\be
\nn\\
&&\hspace{20pt}
-\Big((gg_1\swsw{-}gg_2\sw\cw{-}gg_3\cwcw)\mw^{-2}-g^2\La_V^{-2}\swsw\Big)
[(r+s)_\al p_\be+q_\al(r+s)_\be]
\nn\\
&&\hspace{20pt}
+(g_4+g_6)\oocwcw\mw^{-2}(r_\al s_\be+s_\al r_\be)
\Big\}
\\
\bpi(80,40)(-20,10)
\multiput(0,2)(4,4){4}{\oval(4,4)[br]}
\multiput(4,2)(4,4){4}{\oval(4,4)[tl]}
\multiput(16,14)(4,-4){4}{\oval(4,4)[tr]}
\multiput(20,14)(4,-4){4}{\oval(4,4)[bl]}
\put(0,32){\line(1,-1){16}}
\put(16,16){\line(1,1){16}}
\put(16,16){\circle*{3}}
\put(-20,40){$v_3,r$}
\put(25,40){$v_3,s$}
\put(-22,-12){$Z,p,\al$}
\put(27,-12){$Z,q,\be$}
\epi
&=&
-2i\cw^{-2}\mw^{-2}(g_4+g_5+2g_6+2g_7+2g_8)
(r\cdot sg_{\al\be}+r_\al s_\be+r_\be s_\al)
\\&&\nn\\&&\nn\\
\bpi(80,40)(-20,10)
\multiput(0,2)(4,4){4}{\oval(4,4)[br]}
\multiput(4,2)(4,4){4}{\oval(4,4)[tl]}
\multiput(16,14)(4,-4){4}{\oval(4,4)[tr]}
\multiput(20,14)(4,-4){4}{\oval(4,4)[bl]}
\put(0,32){\line(1,-1){16}}
\put(16,16){\line(1,1){16}}
\put(16,16){\circle*{3}}
\put(-20,40){$v_+,r$}
\put(25,40){$v_-,s$}
\put(-22,-12){$Z,p,\al$}
\put(27,-12){$A,q,\be$}
\epi
&=&
-i\tw\Big\{g_{\al\be}\Big[g^2(\cwcw-\swsw)(1+\La_V^{-2}(p\cdot q)
+2\La_V^{-2}(r\cdot s))
\nn\\
&&\hspace{45pt}
+(gg_1\mw^{-2}+g^2\La_V^{-2})(p\cwcw-q\swsw)\cdot(r+s)
\nn\\
&&\hspace{45pt}
+gg_2\mw^{-2}(p\sw\cw-q\cwcwcw\oosw)\cdot(r+s)
\nn\\
&&\hspace{45pt}
-gg_3\mw^{-2}\cwcw(r+s)^2
\Big]
\nn\\
&&\hspace{25pt}
-g^2\La_V^{-2}(\cwcw-\swsw)(r-s)_\al(r-s)_\be
\nn\\
&&\hspace{25pt}
-\Big((gg_1\cwcw+gg_2\sw\cw+gg_3\cwcw)\mw^{-2}
+g^2\La_V^{-2}\swsw\Big)(r+s)_\al p_\be
\nn\\
&&\hspace{25pt}
+\Big((gg_1\swsw{+}gg_2\cwcwcw\oosw{-}gg_3\cwcw)\mw^{-2}
{+}g^2\La_V^{-2}\cwcw\Big)q_\al(r+s)_\be
\Big\}
\\
\bpi(80,40)(-20,10)
\multiput(0,2)(4,4){4}{\oval(4,4)[br]}
\multiput(4,2)(4,4){4}{\oval(4,4)[tl]}
\multiput(16,14)(4,-4){4}{\oval(4,4)[tr]}
\multiput(20,14)(4,-4){4}{\oval(4,4)[bl]}
\put(0,32){\line(1,-1){16}}
\put(16,16){\line(1,1){16}}
\put(16,16){\circle*{3}}
\put(-20,40){$v_+,r$}
\put(25,40){$v_-,s$}
\put(-22,-12){$A,p,\al$}
\put(27,-12){$A,q,\be$}
\epi
&=&
i\Big\{g_{\al\be}
\Big[2g^2\swsw\left(1+\La_V^{-2}[(p\cdot q)+2(r\cdot s)]\right)
\nn\\
&&\hspace{35pt}
-\Big((gg_1\swsw-gg_2\sw\cw+gg_3\swsw)\mw^{-2}
+g^2\swsw\La_V^{-2}\Big)(p+q)^2\Big]
\nn\\
&&\hspace{15pt}
-2g^2\swsw\La_V^{-2}(r-s)_\al(r-s)_\be
\nn\\
&&\hspace{15pt}
-\Big((gg_1\swsw-gg_2\sw\cw+gg_3\swsw)\mw^{-2}
-g^2\swsw\La_V^{-2}\Big)
[(r+s)_\al p_\be+q_\al(r+s)_\be]
\Big\}
\nn\\
\eea

\subsubsctn{Three-Vertices}
\bea
\bpi(80,40)(-20,10)
\multiput(0,2)(4,4){4}{\oval(4,4)[br]}
\multiput(4,2)(4,4){4}{\oval(4,4)[tl]}
\multiput(16,14)(4,-4){4}{\oval(4,4)[tr]}
\multiput(20,14)(4,-4){4}{\oval(4,4)[bl]}
\multiput(16,18)(0,8){3}{\oval(4,4)[l]}
\multiput(16,22)(0,8){2}{\oval(4,4)[r]}
\put(16,16){\circle*{3}}
\put(0,40){$Z,r,\ga$}
\put(-25,-12){$W^+,p,\al$}
\put(25,-12){$W^-,q,\be$}
\epi
&=&
-i\Big\{\hspace{8pt}
g_{\al\be}\Big[g(\cw+\La_V^{-2}\mw^2\cw)(p-q)_\ga
+g_1\oocw(p-q)_\ga
\nn\\
&&\hspace{50pt}
-g\La_W^{-2}\cw\Big([(p-q)\cdot p]p_\ga+[(p-q)\cdot q]q_\ga\Big)\Big]
\nn\\
&&\hspace{18pt}
+g_{\be\ga}\Big[g(\cw+\La_V^{-2}\mw^2\oocw)(q-r)_\al
+g_1(\oocw q_\al-\cw r_\al)
-(g_2\sw+g_3\cw)r_\al
\nn\\
&&\hspace{50pt}
-g\La_W^{-2}\cw\Big([(q-r)\cdot q]q_\al+[(q-r)\cdot r]r_\al\Big)\Big]
\nn\\
&&\hspace{18pt}
+g_{\al\ga}\Big[g(\cw+\La_V^{-2}\mw^2\oocw)(r-p)_\be
+g_1(\cw r_\be-\oocw p_\be)
+(g_2\sw+g_3\cw)r_\be
\nn\\
&&\hspace{50pt}
-g\La_W^{-2}\cw\Big([(r-p)\cdot r]r_\be+[(r-p)\cdot p]p_\be\Big)\Big]
\nn\\
&&\hspace{18pt}
-g\La_W^{-2}\cw[(p-q)_\ga r_\al r_\be
+(q-r)_\al p_\be p_\ga+(r-p)_\be q_\ga q_\al]
\Big\}
\\&&\nn\\
\bpi(80,40)(-20,10)
\multiput(0,2)(4,4){4}{\oval(4,4)[br]}
\multiput(4,2)(4,4){4}{\oval(4,4)[tl]}
\multiput(16,14)(4,-4){4}{\oval(4,4)[tr]}
\multiput(20,14)(4,-4){4}{\oval(4,4)[bl]}
\multiput(16,18)(0,8){3}{\oval(4,4)[l]}
\multiput(16,22)(0,8){2}{\oval(4,4)[r]}
\put(16,16){\circle*{3}}
\put(0,40){$A,r,\ga$}
\put(-25,-12){$W^+,p,\al$}
\put(25,-12){$W^-,q,\be$}
\epi
&=&
i\sw\Big\{\hspace{5pt}
g_{\al\be}g\Big[(1+\La_V^{-2}\mw^2)(p-q)_\ga
-\La_W^{-2}\Big([(p-q)\cdot p]p_\ga+[(p-q)\cdot q]q_\ga\Big)\Big]
\nn\\
&&\hspace{15pt}
+g_{\be\ga}\Big[g(q{-}r)_\al-(g_1{-}\cw\oosw g_2{+}g_3)r_\al
-g\La_W^{-2}\Big([(q{-}r)\cdot q]q_\al+[(q{-}r)\cdot r]r_\al\Big)\Big]
\nn\\
&&\hspace{15pt}
+g_{\al\ga}\Big[g(r{-}p)_\be+(g_1{-}\cw\oosw g_2{+}g_3)r_\be
-g\La_W^{-2}\Big([(r{-}p)\cdot r]r_\be+[(r{-}p)\cdot p]p_\be\Big)\Big]
\nn\\
&&\hspace{15pt}
-g\La_W^{-2}[(q-r)_\al p_\be p_\ga+(r-p)_\be q_\ga q_\al
+(p-q)_\ga r_\al r_\be]
\Big\}
\\&&\nn\\
\bpi(80,40)(-20,10)
\multiput(0,2)(4,4){4}{\oval(4,4)[br]}
\multiput(4,2)(4,4){4}{\oval(4,4)[tl]}
\multiput(16,14)(4,-4){4}{\oval(4,4)[tr]}
\multiput(20,14)(4,-4){4}{\oval(4,4)[bl]}
\put(16,16){\line(0,1){20}}
\put(16,16){\circle*{3}}
\put(5,40){$v_3,r$}
\put(-30,-12){$W^+,p,\al$}
\put(25,-12){$W^-,q,\be$}
\epi
&=&
\mw^{-1}\left[g_1[(p-q)\cdot r]g_{\al\be}
-\left(g_1+g\La_V^{-2}\mw^2\right)(p_\be r_\al-q_\al r_\be)\right]
\\&&\nn\\&&\nn\\
\bpi(80,40)(-20,10)
\multiput(0,2)(4,4){4}{\oval(4,4)[br]}
\multiput(4,2)(4,4){4}{\oval(4,4)[tl]}
\multiput(16,14)(4,-4){4}{\oval(4,4)[tr]}
\multiput(20,14)(4,-4){4}{\oval(4,4)[bl]}
\put(16,16){\line(0,1){20}}
\put(16,16){\circle*{3}}
\put(0,40){$v_+,r$}
\put(-25,-12){$W^-,p,\al$}
\put(25,-12){$Z,q,\be$}
\epi
&=&-
\bpi(80,40)(-20,10)
\multiput(0,2)(4,4){4}{\oval(4,4)[br]}
\multiput(4,2)(4,4){4}{\oval(4,4)[tl]}
\multiput(16,14)(4,-4){4}{\oval(4,4)[tr]}
\multiput(20,14)(4,-4){4}{\oval(4,4)[bl]}
\put(16,16){\line(0,1){20}}
\put(16,16){\circle*{3}}
\put(0,40){$v_-,r$}
\put(-25,-12){$W^+,p,\al$}
\put(25,-12){$Z,q,\be$}
\epi
\nn\\&&\nn\\&&\nn\\
&=&
\mw\bigg\{g_{\al\be}\bigg[
g\swsw\oocw\left(1{-}\La_V^{-2}p^2\right)
{+}\mw^{-2}g_1\oocw(p\cdot r)
{-}\mw^{-2}\left(g_1\cw{+}g_2\sw{+}g_3\cw\right)(q\cdot r)
\bigg]
\nn\\
&&\hspace{30pt}
-p_\be r_\al\left(g_1\mw^{-2}\oocw+g\La_V^{-2}\cw\right)
+q_\al r_\be\left[\mw^{-2}(g_1\cw{+}g_2\sw{+}g_3\cw)
+g\La_V^{-2}\oocw\right]
\nn\\
&&\hspace{30pt}
-r_\al r_\be g\La_V^{-2}\swsw\oocw\bigg\}
\\&&\nn\\
\bpi(80,40)(-20,10)
\multiput(0,2)(4,4){4}{\oval(4,4)[br]}
\multiput(4,2)(4,4){4}{\oval(4,4)[tl]}
\multiput(16,14)(4,-4){4}{\oval(4,4)[tr]}
\multiput(20,14)(4,-4){4}{\oval(4,4)[bl]}
\put(16,16){\line(0,1){20}}
\put(16,16){\circle*{3}}
\put(0,40){$v_+,r$}
\put(-25,-12){$W^-,p,\al$}
\put(25,-12){$A,q,\be$}
\epi
&=&
-\bpi(80,40)(-20,10)
\multiput(0,2)(4,4){4}{\oval(4,4)[br]}
\multiput(4,2)(4,4){4}{\oval(4,4)[tl]}
\multiput(16,14)(4,-4){4}{\oval(4,4)[tr]}
\multiput(20,14)(4,-4){4}{\oval(4,4)[bl]}
\put(16,16){\line(0,1){20}}
\put(16,16){\circle*{3}}
\put(0,40){$v_-,r$}
\put(-25,-12){$W^+,p,\al$}
\put(25,-12){$A,q,\be$}
\epi
\nn\\&&\nn\\&&\nn\\
&=&
\mw\bigg\{\left[g\sw\left(1{-}\La_V^{-2}p^2\right)
{+}\mw^{-2}(g_1\sw{-}g_2\cw{+}g_3\sw)
(q\cdot r)\right]g_{\al\be}
\nn\\
&&\hspace{30pt}
+g\sw\La_V^{-2}r_\al(p{-}r)_\be
+\mw^{-2}(-g_1\sw{+}g_2\cw{-}g_3\sw)q_\al r_\be\bigg\}
\nn\\&&\\
\bpi(80,40)(-20,10)
\put(0,0){\line(1,1){16}}
\put(16,16){\line(1,-1){16}}
\multiput(16,18)(0,8){3}{\oval(4,4)[l]}
\multiput(16,22)(0,8){2}{\oval(4,4)[r]}
\put(16,16){\circle*{3}}
\put(-10,40){$W^-,p,\al$}
\put(-17,-12){$v_+,q$}
\put(27,-12){$v_3,r$}
\epi
&=&
-\bpi(80,40)(-20,10)
\put(0,0){\line(1,1){16}}
\put(16,16){\line(1,-1){16}}
\multiput(16,18)(0,8){3}{\oval(4,4)[l]}
\multiput(16,22)(0,8){2}{\oval(4,4)[r]}
\put(16,16){\circle*{3}}
\put(-10,40){$W^+,p,\al$}
\put(-17,-12){$v_-,q$}
\put(27,-12){$v_3,r$}
\epi
\nn\\&&\nn\\&&\nn\\
&=&
-i\Big\{\half g\left[1-\La_V^{-2}\left(q^2+r^2\right)\right](q-r)_\al
-g_1\mw^{-2}[(p\cdot q)r_\al-(p\cdot r)q_\al]\Big\}
\\
\bpi(80,40)(-20,10)
\put(0,0){\line(1,1){16}}
\put(16,16){\line(1,-1){16}}
\multiput(16,18)(0,8){3}{\oval(4,4)[l]}
\multiput(16,22)(0,8){2}{\oval(4,4)[r]}
\put(16,16){\circle*{3}}
\put(0,40){$Z,p,\al$}
\put(-17,-12){$v_+,q$}
\put(27,-12){$v_-,r$}
\epi
&=&
-i\Big\{\mw^{-2}
(g_1\cw+g_2\sw+g_3\cw)[(p\cdot q)r_\al-(p\cdot r)q_\al]
\nn\\
&&\hspace{20pt}
-\half g\oocw\left[
(\cwcw-\swsw)\left(1+\La_V^{-2}[(r-q)\cdot q]\right)
+\La_V^{-2}(p\cdot r)\right]q_\al
\nn\\
&&\hspace{20pt}
+\half g\oocw\left[
(\cwcw-\swsw)\left(1+\La_V^{-2}[(q-r)\cdot r]\right)
+\La_V^{-2}(p\cdot q)\right]r_\al
\Big\}
\nn\\&&\\
\bpi(80,40)(-20,10)
\put(0,0){\line(1,1){16}}
\put(16,16){\line(1,-1){16}}
\multiput(16,18)(0,8){3}{\oval(4,4)[l]}
\multiput(16,22)(0,8){2}{\oval(4,4)[r]}
\put(16,16){\circle*{3}}
\put(0,40){$A,p,\al$}
\put(-17,-12){$v_+,q$}
\put(27,-12){$v_-,r$}
\epi
&=&
i\Big\{\mw^{-2}(g_1\sw-g_2\cw+g_3\sw)[(p\cdot q)r_\al-(p\cdot r)q_\al]
\nn\\
&&\hspace{10pt}
-g\sw\left[\left(1+\La_V^{-2}[(r-q)\cdot q]\right)q_\al
-\left(1+\La_V^{-2}[(q-r)\cdot r]\right)r_\al\right]\Big\}
\\&&\nn\\&&\nn\\
\bpi(80,40)(-20,10)
\multiput(0,0)(3,3){6}{\circle*{1}}
\multiput(32,0)(-3,3){6}{\circle*{1}}
\multiput(16,18)(0,8){3}{\oval(4,4)[l]}
\multiput(16,22)(0,8){2}{\oval(4,4)[r]}
\put(23.5,4.5){\line(1,0){4}}
\put(27.5,4.5){\line(0,1){4}}
\put(3.5,7.5){\line(1,0){4}}
\put(7.5,3.5){\line(0,1){4}}
\put(16,16){\circle*{3}}
\put(-10,40){$W^-,p,\al$}
\put(-13,-12){$\et_{\sss W+},r$}
\put(27,-12){$\eb_{\sss Z},q$}
\epi
&=&-
\bpi(80,40)(-20,10)
\multiput(0,0)(3,3){6}{\circle*{1}}
\multiput(32,0)(-3,3){6}{\circle*{1}}
\multiput(16,18)(0,8){3}{\oval(4,4)[l]}
\multiput(16,22)(0,8){2}{\oval(4,4)[r]}
\put(23.5,4.5){\line(1,0){4}}
\put(27.5,4.5){\line(0,1){4}}
\put(3.5,7.5){\line(1,0){4}}
\put(7.5,3.5){\line(0,1){4}}
\put(16,16){\circle*{3}}
\put(-10,40){$W^+,p,\al$}
\put(-13,-12){$\et_{\sss W-},r$}
\put(27,-12){$\eb_{\sss Z},q$}
\epi
=
ig\cw q_\al
\\&&\nn\\&&\nn\\
\bpi(80,40)(-20,10)
\multiput(0,0)(3,3){6}{\circle*{1}}
\multiput(32,0)(-3,3){6}{\circle*{1}}
\multiput(16,18)(0,8){3}{\oval(4,4)[l]}
\multiput(16,22)(0,8){2}{\oval(4,4)[r]}
\put(23.5,4.5){\line(1,0){4}}
\put(27.5,4.5){\line(0,1){4}}
\put(3.5,7.5){\line(1,0){4}}
\put(7.5,3.5){\line(0,1){4}}
\put(16,16){\circle*{3}}
\put(-10,40){$W^-,p,\al$}
\put(-13,-12){$\et_{\sss Z},r$}
\put(27,-12){$\eb_{\sss W+},q$}
\epi
&=&-
\bpi(80,40)(-20,10)
\multiput(0,0)(3,3){6}{\circle*{1}}
\multiput(32,0)(-3,3){6}{\circle*{1}}
\multiput(16,18)(0,8){3}{\oval(4,4)[l]}
\multiput(16,22)(0,8){2}{\oval(4,4)[r]}
\put(23.5,4.5){\line(1,0){4}}
\put(27.5,4.5){\line(0,1){4}}
\put(3.5,7.5){\line(1,0){4}}
\put(7.5,3.5){\line(0,1){4}}
\put(16,16){\circle*{3}}
\put(-10,40){$W^+,p,\al$}
\put(-13,-12){$\et_{\sss Z},r$}
\put(27,-12){$\eb_{\sss W-},q$}
\epi
=
-ig\cw q_\al
\\&&\nn\\&&\nn\\
\bpi(80,40)(-20,10)
\multiput(0,0)(3,3){6}{\circle*{1}}
\multiput(32,0)(-3,3){6}{\circle*{1}}
\multiput(16,18)(0,8){3}{\oval(4,4)[l]}
\multiput(16,22)(0,8){2}{\oval(4,4)[r]}
\put(23.5,4.5){\line(1,0){4}}
\put(27.5,4.5){\line(0,1){4}}
\put(3.5,7.5){\line(1,0){4}}
\put(7.5,3.5){\line(0,1){4}}
\put(16,16){\circle*{3}}
\put(0,40){$Z,p,\al$}
\put(-13,-12){$\et_{\sss W+},r$}
\put(27,-12){$\eb_{\sss W-},q$}
\epi
&=&-
\bpi(80,40)(-20,10)
\multiput(0,0)(3,3){6}{\circle*{1}}
\multiput(32,0)(-3,3){6}{\circle*{1}}
\multiput(16,18)(0,8){3}{\oval(4,4)[l]}
\multiput(16,22)(0,8){2}{\oval(4,4)[r]}
\put(23.5,4.5){\line(1,0){4}}
\put(27.5,4.5){\line(0,1){4}}
\put(3.5,7.5){\line(1,0){4}}
\put(7.5,3.5){\line(0,1){4}}
\put(16,16){\circle*{3}}
\put(0,40){$Z,p,\al$}
\put(-13,-12){$\et_{\sss W-},r$}
\put(27,-12){$\eb_{\sss W+},q$}
\epi
=
-ig\cw q_\al
\\&&\nn\\&&\nn\\
\bpi(80,40)(-20,10)
\multiput(0,0)(3,3){6}{\circle*{1}}
\multiput(32,0)(-3,3){6}{\circle*{1}}
\multiput(16,18)(0,8){3}{\oval(4,4)[l]}
\multiput(16,22)(0,8){2}{\oval(4,4)[r]}
\put(23.5,4.5){\line(1,0){4}}
\put(27.5,4.5){\line(0,1){4}}
\put(3.5,7.5){\line(1,0){4}}
\put(7.5,3.5){\line(0,1){4}}
\put(16,16){\circle*{3}}
\put(-10,40){$W^-,p,\al$}
\put(-13,-12){$\et_{\sss W+},r$}
\put(27,-12){$\eb_{\sss A},q$}
\epi
&=&-
\bpi(80,40)(-20,10)
\multiput(0,0)(3,3){6}{\circle*{1}}
\multiput(32,0)(-3,3){6}{\circle*{1}}
\multiput(16,18)(0,8){3}{\oval(4,4)[l]}
\multiput(16,22)(0,8){2}{\oval(4,4)[r]}
\put(23.5,4.5){\line(1,0){4}}
\put(27.5,4.5){\line(0,1){4}}
\put(3.5,7.5){\line(1,0){4}}
\put(7.5,3.5){\line(0,1){4}}
\put(16,16){\circle*{3}}
\put(-10,40){$W^+,p,\al$}
\put(-13,-12){$\et_{\sss W-},r$}
\put(27,-12){$\eb_{\sss A},q$}
\epi
=
-ig\sw q_\al
\\&&\nn\\&&\nn\\
\bpi(80,40)(-20,10)
\multiput(0,0)(3,3){6}{\circle*{1}}
\multiput(32,0)(-3,3){6}{\circle*{1}}
\multiput(16,18)(0,8){3}{\oval(4,4)[l]}
\multiput(16,22)(0,8){2}{\oval(4,4)[r]}
\put(23.5,4.5){\line(1,0){4}}
\put(27.5,4.5){\line(0,1){4}}
\put(3.5,7.5){\line(1,0){4}}
\put(7.5,3.5){\line(0,1){4}}
\put(16,16){\circle*{3}}
\put(-10,40){$W^-,p,\al$}
\put(-13,-12){$\et_{\sss A},r$}
\put(27,-12){$\eb_{\sss W+},q$}
\epi
&=&-
\bpi(80,40)(-20,10)
\multiput(0,0)(3,3){6}{\circle*{1}}
\multiput(32,0)(-3,3){6}{\circle*{1}}
\multiput(16,18)(0,8){3}{\oval(4,4)[l]}
\multiput(16,22)(0,8){2}{\oval(4,4)[r]}
\put(23.5,4.5){\line(1,0){4}}
\put(27.5,4.5){\line(0,1){4}}
\put(3.5,7.5){\line(1,0){4}}
\put(7.5,3.5){\line(0,1){4}}
\put(16,16){\circle*{3}}
\put(-10,40){$W^+,p,\al$}
\put(-13,-12){$\et_{\sss A},r$}
\put(27,-12){$\eb_{\sss W-},q$}
\epi
=
ig\sw q_\al
\\&&\nn\\&&\nn\\
\bpi(80,40)(-20,10)
\multiput(0,0)(3,3){6}{\circle*{1}}
\multiput(32,0)(-3,3){6}{\circle*{1}}
\multiput(16,18)(0,8){3}{\oval(4,4)[l]}
\multiput(16,22)(0,8){2}{\oval(4,4)[r]}
\put(23.5,4.5){\line(1,0){4}}
\put(27.5,4.5){\line(0,1){4}}
\put(3.5,7.5){\line(1,0){4}}
\put(7.5,3.5){\line(0,1){4}}
\put(16,16){\circle*{3}}
\put(0,40){$A,p,\al$}
\put(-13,-12){$\et_{\sss W+},r$}
\put(27,-12){$\eb_{\sss W-},q$}
\epi
&=&-
\bpi(80,40)(-20,10)
\multiput(0,0)(3,3){6}{\circle*{1}}
\multiput(32,0)(-3,3){6}{\circle*{1}}
\multiput(16,18)(0,8){3}{\oval(4,4)[l]}
\multiput(16,22)(0,8){2}{\oval(4,4)[r]}
\put(23.5,4.5){\line(1,0){4}}
\put(27.5,4.5){\line(0,1){4}}
\put(3.5,7.5){\line(1,0){4}}
\put(7.5,3.5){\line(0,1){4}}
\put(16,16){\circle*{3}}
\put(0,40){$A,p,\al$}
\put(-13,-12){$\et_{\sss W-},r$}
\put(27,-12){$\eb_{\sss W+},q$}
\epi
=
ig\sw q_\al
\\&&\nn\\&&\nn\\
\bpi(80,40)(-20,10)
\multiput(0,0)(3,3){6}{\circle*{1}}
\multiput(32,0)(-3,3){6}{\circle*{1}}
\put(16,16){\line(0,1){20}}
\put(23.5,4.5){\line(1,0){4}}
\put(27.5,4.5){\line(0,1){4}}
\put(3.5,7.5){\line(1,0){4}}
\put(7.5,3.5){\line(0,1){4}}
\put(16,16){\circle*{3}}
\put(5,40){$v_3,p$}
\put(-13,-12){$\et_{\sss W-},r$}
\put(27,-12){$\eb_{\sss W+},q$}
\epi
&=&-
\bpi(80,40)(-20,10)
\multiput(0,0)(3,3){6}{\circle*{1}}
\multiput(32,0)(-3,3){6}{\circle*{1}}
\put(16,16){\line(0,1){20}}
\put(23.5,4.5){\line(1,0){4}}
\put(27.5,4.5){\line(0,1){4}}
\put(3.5,7.5){\line(1,0){4}}
\put(7.5,3.5){\line(0,1){4}}
\put(16,16){\circle*{3}}
\put(5,40){$v_3,p$}
\put(-13,-12){$\et_{\sss W+},r$}
\put(27,-12){$\eb_{\sss W-},q$}
\epi
=
\half \xi g\mw(1-\La_V^{-2}q^2)
\eea
\mbox{\ }

\sctn{One-loop Integrals}
\label{integrals}

Define $\ep$ by
\beq
d=4-2\ep\,,
\eeq
where $d$ is the spacetime dimension, and $\mubar$ by
\beq
\ln4\pi\mu^2-\ga_{\sss E}=\ln\mubar^2
\eeq
and $\int_p$ by
\beq
\int_p=\int\frac{d^dp}{(2\pi)^d}\,.
\eeq

The only integrals we need are
\beq
I(m^2)\equiv\int_p\frac{1}{p^2-m^2}=
\frac{im^2}{(4\pi)^2}\left(\frac{1}{\ep}+1-\ln\frac{m^2}{\mubar^2}\right)
+\od(\ep)
\eeq
and 
\bea
\label{ikmambdef}
I(k^2;m_a^2,m_b^2)
&\equiv&
\int_p\frac{1}{[(p+k)^2-m_a^2+i\ve](p^2-m_b^2+i\ve)}
\nn\\
&=&
\int_0^1dx\int_p\frac{1}{[p^2+2xp\cdot k+xk^2-xm_a^2-(1-x)m_b^2+i\ve]^2}
\nn\\
&=&
\frac{i\mu^{2\ep}\Ga(\ep)}{(4\pi)^{2-\ep}}
\int_0^1\frac{dx}{[-x(1-x)k^2+xm_a^2+(1-x)m_b^2-i\ve]^\ep}
\nn\\
&=&
\frac{i}{(4\pi)^2}\left(\frac{1}{\ep}-\ga_{\sss E}\right)
\left(1-\ep\int_0^1dx\ln
\frac{-x(1-x)k^2+xm_a^2+(1-x)m_b^2-i\ve}{4\pi\mu^2}\right)
+\od(\ep)
\nn\\
\label{ikmamb}
&=&
\frac{i}{(4\pi)^2}\left(\frac{1}{\ep}-\int_0^1dx\ln
\frac{-x(1-x)k^2+xm_a^2+(1-x)m_b^2-i\ve}{\mubar^2}\right)
+\od(\ep)
\nn\\
&=&
\frac{i}{(4\pi)^2}\left(\frac{1}{\ep}-\int_0^1dx\ln
\frac{k^2(x-x_0)^2-D/(4k^2)-i\ve}{\mubar^2}\right)
+\od(\ep)\,,
\eea
where
\beq
x_0\equiv\frac{k^2+m_b^2-m_a^2}{2k^2}
\eeq
and
\beq
D\equiv k^4+m_a^4+m_b^4-2k^2m_a^2-2k^2m_b^2-2m_a^2m_b^2\,.
\eeq
We need to investigate here only the case where the argument of the
logarithm is non-negative for $0\leq x\leq 1$ and therefore
$I(k^2;m_a^2,m_b^2)$ is purely imaginary.
This is obviously the case for $D\leq0$.
For $D>0$ this is the case if and only if $x_0\leq0$ or $x_0\geq1$,
i.e.\ $k^2\leq |m_a^2-m_b^2|$.

\subsctn{$I(k^2;m_a^2,m_b^2)$ for $D\leq0$}
Now we can write
\bea
\label{ikmm1}
I(k^2;m_a^2,m_b^2)
&=&
\frac{i}{(4\pi)^2}\left(\frac{1}{\ep}-\ln\frac{k^2}{\mubar^2}
-\int_{-x_0}^{1-x_0}dy\ln\left(y^2+\frac{-D}{4k^4}\right)\right)
+\od(\ep)
\nn\\
&=&
\frac{i}{(4\pi)^2}\left\{\frac{1}{\ep}-\ln\frac{k^2}{\mubar^2}
-\left[y\ln\left(y^2+\frac{-D}{4k^4}\right)
-2y+2\sqrt{\frac{-D}{4k^4}}
\arctan\frac{y}{\sqrt{\frac{-D}{4k^4}}}
\right]_{-x_0}^{1-x_0}
\right\}+\od(\ep)
\nn\\
&=&
\frac{i}{(4\pi)^2}\Bigg[\frac{1}{\ep}+2-\ln\frac{k^2}{\mubar^2}
-\frac{k^2+m_a^2-m_b^2}{2k^2}\ln\frac{m_a^2}{k^2}
-\frac{k^2+m_b^2-m_a^2}{2k^2}\ln\frac{m_b^2}{k^2}
\nn\\
&&\hspace{35pt}
-\frac{\sqrt{-D}}{k^2}\left(\arctan\frac{k^2+m_a^2-m_b^2}{\sqrt{-D}}
+\arctan\frac{k^2+m_b^2-m_a^2}{\sqrt{-D}}\right)
\Bigg]+\od(\ep)
\nn\\
&=&
\frac{i}{(4\pi)^2}\Bigg[\frac{1}{\ep}+2
-\frac{k^2+m_a^2-m_b^2}{2k^2}\ln\frac{m_a^2}{\mubar^2}
-\frac{k^2+m_b^2-m_a^2}{2k^2}\ln\frac{m_b^2}{\mubar^2}
\nn\\
&&\hspace{35pt}
-\frac{\sqrt{-D}}{k^2}\left(\arctan\frac{k^2+m_a^2-m_b^2}{\sqrt{-D}}
+\arctan\frac{k^2+m_b^2-m_a^2}{\sqrt{-D}}\right)
\Bigg]+\od(\ep)\,.
\eea

\subsctn{$I(k^2;m_a^2,m_b^2)$ for $D\geq0$ with $k^2\leq |m_a^2-m_b^2|$}
Define
\beq
x_\pm\equiv\frac{k^2+m_b^2-m_a^2\pm\sqrt{D}}{2k^2}\,,
\eeq
so that
\beq
1-x_\pm=\frac{k^2+m_a^2-m_b^2\mp\sqrt{D}}{2k^2}\,.
\eeq
Without loss of generality assume $m_a^2\geq m_b^2$.
Then $x_\pm\leq 0$ and $1-x_\pm\geq0$.
We can write
\bea
\label{ikmm2}
I(k^2;m_a^2,m_b^2)
&=&
\frac{i}{(4\pi)^2}\left\{\frac{1}{\ep}
-\ln\frac{k^2}{\mubar^2}
-\int_0^1dx\ln[(x-x_+)(x-x_-)]\right\}
+\od(\ep)
\nn\\
&=&
\frac{i}{(4\pi)^2}\left[\frac{1}{\ep}
-\ln\frac{k^2}{\mubar^2}
-\int_0^1dx\ln(x-x_+)-\int_0^1dx\ln(x-x_-)\right]
+\od(\ep)
\nn\\
&=&
\frac{i}{(4\pi)^2}\bigg\{\frac{1}{\ep}-\ln\frac{k^2}{\mubar^2}
-(1-x_+)[\ln(1-x_+)-1]-x_+[\ln(-x_+)-1]
\nn\\
&&\hspace{90pt}
-(1-x_-)[\ln(1-x_-)-1]-x_-[\ln(-x_-)-1]\bigg\}+\od(\ep)
\nn\\
&=&
\frac{i}{(4\pi)^2}\bigg[\frac{1}{\ep}+2-\ln\frac{k^2}{\mubar^2}
-(1-x_+)\ln(1-x_+)-x_+\ln(-x_+)
\nn\\
&&\hspace{110pt}
-(1-x_-)\ln(1-x_-)-x_-\ln(-x_-)\bigg]+\od(\ep)\,.
\eea
Now one can write either
\bea
\label{ikmm2a}
\lefteqn{I(k^2;m_a^2,m_b^2)}
\nn\\
&=&
\frac{i}{(4\pi)^2}\bigg\{\frac{1}{\ep}+2-\ln\frac{k^2}{\mubar^2}
-x_+\ln[(-x_+)(-x_-)]-(1-x_-)\ln[(1-x_+)(1-x_-)]
\nn\\
&&\hspace{35pt}
+(x_+-x_-)\ln[(1-x_+)(-x_-)]\bigg\}+\od(\ep)
\nn\\
&=&
\frac{i}{(4\pi)^2}\bigg\{\frac{1}{\ep}+2
-\frac{k^2-m_a^2+m_b^2+\sqrt{D}}{2k^2}
\ln\frac{m_b^2}{\mubar^2}
-\frac{k^2+m_a^2-m_b^2+\sqrt{D}}{2k^2}
\ln\frac{m_a^2}{\mubar^2}
\nn\\
&&\hspace{35pt}
+\frac{\sqrt{D}}{k^2}
\ln\frac{m_a^2+m_b^2-k^2+\sqrt{D}}{2\mubar^2}
\bigg\}+\od(\ep)
\nn\\
\eea
or
\bea
\label{ikmm2b}
\lefteqn{I(k^2;m_a^2,m_b^2)}
\nn\\
&=&
\frac{i}{(4\pi)^2}\bigg\{\frac{1}{\ep}+2-\ln\frac{k^2}{\mubar^2}
-x_-\ln[(-x_-)(-x_+)]-(1-x_+)\ln[(1-x_-)(1-x_+)]
\nn\\
&&\hspace{35pt}
+(x_--x_+)\ln[(1-x_-)(-x_+)]\bigg\}+\od(\ep)
\nn\\
&=&
\frac{i}{(4\pi)^2}\bigg\{\frac{1}{\ep}+2
-\frac{k^2-m_a^2+m_b^2-\sqrt{D}}{2k^2}
\ln\frac{m_b^2}{\mubar^2}
-\frac{k^2+m_a^2-m_b^2-\sqrt{D}}{2k^2}
\ln\frac{m_a^2}{\mubar^2}
\nn\\
&&\hspace{35pt}
-\frac{\sqrt{D}}{k^2}
\ln\frac{m_a^2+m_b^2-k^2-\sqrt{D}}{2\mubar^2}
\bigg\}+\od(\ep)
\nn\\
\eea
with
\beq
\sqrt{D}\equiv\sqrt{k^4+m_a^4+m_b^4-2m_a^2m_b^2-2k^2m_a^2-2k^2m_b^2}\,.
\eeq
(\ref{ikmm2a}) and (\ref{ikmm2b}) are symmetric in $m_a^2$ and $m_b^2$
and therefore we can drop the restriction $m_a^2\geq m_b^2$.

In the following we will specialize to the cases that are needed for the
evaluation of our one-loop diagrams.

\subsctn{$I(k^2;m^2,m^2)$}
Only for $D=k^2(k^2-4m^2)\leq0$, i.e.\ for $k^2\leq4m^2$
we have purely imaginary $I(k^2;m^2,m^2)$.
From (\ref{ikmm1}) we get
\beq
\label{ikmama}
I(k^2;m^2,m^2)=
\frac{i}{(4\pi)^2}\left(\frac{1}{\ep}+2-\ln\frac{m^2}{\mubar^2}
-2{\ts\sqrt{\frac{4m^2}{k^2}-1}}\;
\arctan\frac{1}{\sqrt{\frac{4m^2}{k^2}-1}}\right)+\od(\ep)\,.
\eeq

For $k^2\ll m^2$, we can expand in powers of $k^2/m^2$ to get
\beq
I(k^2;m^2,m^2)=
\frac{i}{(4\pi)^2}\left[\frac{1}{\ep}-\ln\frac{m^2}{\mubar^2}
+\frac{1}{6}\left(\frac{k^2}{m^2}\right)
+\frac{1}{60}\left(\frac{k^2}{m^2}\right)^2\right]
+\od\left(\ep,\left(\ts\frac{k^2}{m^2}\right)^3\right)\,.
\eeq

\subsctn{$I(k^2;m^2,0)$}
Now $D=|k^2-m^2|\geq0$ and we need $k^2\leq m^2$ to have a purely
imaginary $I(k^2;m^2,0)$.
We get from (\ref{ikmm2a}) and (\ref{ikmm2b})
\bea
\label{ikm0}
I(k^2;0,m^2)=I(k^2;m^2,0)
&=&
\frac{i}{(4\pi)^2}\left[\frac{1}{\ep}+2
+\frac{m^2-k^2}{k^2}\ln\frac{m^2-k^2}{\mubar^2} 
-\frac{m^2}{k^2}\ln\frac{m^2}{\mubar^2}\right]+\od(\ep)
\nn\\
&=&
\frac{i}{(4\pi)^2}\left[\frac{1}{\ep}+2
+\frac{m^2}{k^2}\ln\left(1-\frac{k^2}{m^2}\right) 
-\ln\frac{m^2-k^2}{\mubar^2}\right]+\od(\ep)\,.
\eea

For $k^2\ll m^2$, we can expand in powers of $k^2/m^2$ to get
\bea
I(k^2;0,m^2)=I(k^2;m^2,0)=
\frac{i}{(4\pi)^2}\left[\frac{1}{\ep}+1-\ln\frac{m^2}{\mubar^2}
+\frac{1}{2}\left(\frac{k^2}{m^2}\right)
+\frac{1}{6}\left(\frac{k^2}{m^2}\right)^2
\right]+\od(\ep,\left(\ts\frac{k^2}{m^2}\right)^3)\,.
\eea

\subsctn{$I(k^2;m_a^2,m_b^2)$ for $k^2,m_a^2\ll m_b^2$}
If $k^2,m_a^2\ll m_b^2$, we can expand (\ref{ikmm2a}) or (\ref{ikmm2b})
in negative powers of $m_b^2$ to get
\bea
I(k^2;m_a^2,m_b^2)
&=&
\frac{i}{(4\pi)^2}
\bigg[\frac{1}{\ep}+1-\ln\frac{m_b^2}{\mubar^2}
+\frac{\frac{1}{2}k^2+m_a^2\ln\frac{m_a^2}{m_b^2}}{m_b^2}
+\frac{k^2(\frac{1}{6}k^2+\frac{3}{2}m_a^2)+
m_a^2(k^2+m_a^2)\ln\frac{m_a^2}{m_b^2}}{m_b^4}
\bigg]
\nn\\
&&
+\od(\ep,m_b^{-6}\ln m_b^2)\,.
\eea

\subsctn{$I(k^2;m_a^2,m_b^2)$ for $k^2\ll m_a^2,m_b^2$}
If $k^2\ll m_a^2,m_b^2$, but the relative magnitude of $k^2$ and
$|m_a^2-m_b^2|$ is unknown, it is not clear, which of (\ref{ikmm1})
on the one hand or (\ref{ikmm2a}), (\ref{ikmm2b}) on the other
hand has to be used.
Although they are connected by analytic continuation, here we will
expand $I(k^2;m_a^2,m_b^2)$ in powers of $k^2$ to have an unambiguous
result without without having to worry about Riemann sheets.

Starting from the next-to-last line in (\ref{ikmambdef}) we get
\bea
\label{ik2mamb}
\lefteqn{I(k^2;m_a^2,m_b^2)}
\nn\\
&=&
\frac{i}{(4\pi)^2}\left(\frac{1}{\ep}-\int_0^1dx\ln
\frac{-x(1-x)k^2+xm_a^2+(1-x)m_b^2}{\mubar^2}\right)
+\od(\ep)
\nn\\
&=&
\frac{i}{(4\pi)^2}\left[\frac{1}{\ep}
-\int_0^1dx\ln\frac{xm_a^2+(1-x)m_b^2}{\mubar^2}
-\int_0^1dx\ln\left(1-\frac{x(1-x)k^2}{xm_a^2+(1-x)m_b^2}\right)
\right]
+\od(\ep)
\nn\\
&=&
\frac{i}{(4\pi)^2}\left[\frac{1}{\ep}+1
-\frac{m_a^2\ln\frac{m_a^2}{\mubar^2}
-m_b^2\ln\frac{m_b^2}{\mubar^2}}{m_a^2-m_b^2}
+\sum_{n=1}^\infty\frac{1}{n}\int_0^1dx
\left(\frac{x(1-x)k^2}{xm_a^2+(1-x)m_b^2}\right)^n\right]
+\od(\ep)\,.
\eea
Expanding in $k^2$, we get
\bea
\label{ik2mambex}
\lefteqn{I(k^2;m_a^2,m_b^2)}
\nn\\
&=&
\frac{i}{(4\pi)^2}\Bigg[\frac{1}{\ep}+1
-\frac{m_a^2\ln\frac{m_a^2}{\mubar^2}
-m_b^2\ln\frac{m_b^2}{\mubar^2}}{m_a^2-m_b^2}
\nn\\
&&\hspace{35pt}
+k^2\int_0^1dx\left(\frac{x(1-x)}{xm_a^2+(1-x)m_b^2}\right)
+\frac{k^4}{2}\int_0^1dx\left(\frac{x(1-x)}{xm_a^2+(1-x)m_b^2}\right)^2
\Bigg]+\od(k^6,\ep)
\nn\\
&=&
\frac{i}{(4\pi)^2}\Bigg[\frac{1}{\ep}+1
-\frac{m_a^2\ln\frac{m_a^2}{\mubar^2}
-m_b^2\ln\frac{m_b^2}{\mubar^2}}{m_a^2-m_b^2}
+\Bigg(\frac{m_a^2+m_b^2}{2(m_a^2-m_b^2)^2}
-\frac{m_a^2m_b^2}{(m_a^2-m_b^2)^3}\ln\frac{m_a^2}{m_b^2}\Bigg)k^2
\nn\\
&&\hspace{35pt}
+\Bigg(\frac{m_a^4+10m_a^2m_b^2+m_b^4}{6(m_a^2-m_b^2)^4}
-\frac{m_a^2m_b^2(m_a^2+m_b^2)}{(m_a^2-m_b^2)^5}
\ln\frac{m_a^2}{m_b^2}\Bigg)k^4\Bigg]+\od(k^6,\ep)\,.
\eea
Note that (\ref{ik2mamb}) tells us that subsequent powers of $k^2$
in (\ref{ik2mambex}) are suppressed by negative powers of $m_a^2$
and $m_b^2$ and not just by their difference $m_a^2-m_b^2$, which
might be small or even vanishing.

Indeed, setting $m_a^2=m^2+\de m_a^2$, $m_b^2=m^2+\de m_b^2$ with
$k^2,\de m_a^2,\de m_b^2\ll m^2$ and starting again from the
next-to-last line in (\ref{ikmambdef}) we get
\bea
\lefteqn{I(k^2;m^2+\de m_a^2,m^2+\de m_b^2)}
\nn\\
&=&
\frac{i}{(4\pi)^2}\left(\frac{1}{\ep}-\int_0^1dx\ln
\frac{m^2-x(1-x)k^2+x\de m_a^2+(1-x)\de m_b^2}{\mubar^2}\right)
+\od(\ep)
\nn\\
&=&
\frac{i}{(4\pi)^2}\left[\frac{1}{\ep}
-\ln\frac{m^2}{\mubar^2}-\int_0^1dx\ln\left(1-
\frac{x(1-x)k^2-x\de m_a^2-(1-x)\de m_b^2}{m^2}\right)\right]
+\od(\ep)
\nn\\
&=&
\frac{i}{(4\pi)^2}\left[\frac{1}{\ep}
-\ln\frac{m^2}{\mubar^2}
+\sum_{n=1}^\infty\frac{1}{n}\int_0^1dx
\left(\frac{x(1-x)k^2-x\de m_a^2-(1-x)\de m_b^2}{m^2}\right)^n
\right]+\od(\ep)
\nn\\
&=&
\frac{i}{(4\pi)^2}\bigg[\frac{1}{\ep}
-\ln\frac{m^2}{\mubar^2}
+\frac{1}{6}\left(\frac{k^2}{m^2}\right)
-\frac{1}{2}\left(\frac{\de m_a^2}{m^2}\right)
-\frac{1}{2}\left(\frac{\de m_b^2}{m^2}\right)
\nn\\
&&\hspace{30pt}
+\frac{1}{60}\left(\frac{k^2}{m^2}\right)^2
+\frac{1}{6}\left(\frac{\de m_a^2}{m^2}\right)^2
+\frac{1}{6}\left(\frac{\de m_b^2}{m^2}\right)^2
\nn\\
&&\hspace{30pt}
-\frac{1}{12}\left(\frac{k^2}{m^2}\right)
\left(\frac{\de m_a^2}{m^2}\right)
-\frac{1}{12}\left(\frac{k^2}{m^2}\right)
\left(\frac{\de m_b^2}{m^2}\right)
+\frac{1}{6}\left(\frac{\de m_a^2}{m^2}\right)
\left(\frac{\de m_b^2}{m^2}\right)
\bigg]
\nn\\
&&
+\od(m^{-6},\ep)\,,
\eea
which can also be obtained by expanding (\ref{ik2mambex}).

\end{document}